\documentclass[pra,aps,amssymb,twocolumn,noshowpacs]{revtex4-1}
\usepackage{color}
\usepackage{graphicx}
\usepackage{hyperref}
\usepackage{amsmath}
\usepackage{latexsym}
\usepackage{amssymb}
\usepackage{mathrsfs}
\usepackage{mathtools}
\usepackage{layout}
\usepackage{verbatim}
\usepackage{multirow}
\usepackage{bm}
\usepackage{amsfonts,epsfig}

\begin{document}
\title{Hyperentanglement of divalent neutral atoms by Rydberg blockade}

\date{\today}
\author{Xiao-Feng Shi}
\affiliation{School of Physics and Optoelectronic Engineering, Xidian University, Xi'an 710071, China}

\begin{abstract}
  Hyperentanglement~(HE), the simultaneous entanglement between two particles in more than one degrees of freedom, is relevant to both fundamental physics and quantum technology. Previous study on HE has been focusing on photons. Here, we study HE in individual neutral atoms. In most alkaline-earth-like atoms with two valence electrons and a nonzero nuclear spin, there are two stable electronic states, the ground state and the long-lived clock state, which can define an electronic qubit. Meanwhile, their nuclear spin states can define a nuclear qubit. By the Rydberg blockade effect, we show that the controlled-Z~(C$_{\text{Z}}$) operation can be generated in the electronic qubits of two nearby atoms, and simultaneously in their nuclear qubits as well, leading to a C$_{\text{Z}}\otimes$C$_{\text{Z}}$ operation which is capable to induce HE. The possibility to induce HE in individual neutral atoms offers new opportunities to study quantum science and technology based on neutral atoms.

\end{abstract}
\maketitle

\section{introduction}\label{sec01}

An exotic multidimensional entanglement phenomenon is hyperentanglement~(HE), namely, a simultaneous entanglement in each of two or more than two degrees of freedom. The capability to entangle two particles in more than one degrees of freedom can enhance the information and quantum correlation carried by the particle pairs which brings extra strength in the investigation of fundamental quantum theory, quantum metrology, and quantum information. HE was extensively studied in photonic systems~\cite{Walborn2003,Cinelli2005,Schuck2006,Barbieri2006,Chen2007,Gao2010,Zhao2019,Chen2020,Chen20202,Graffitti2020,Hu2021}, but much less in other candidates for quantum information.

Recently, with remarkable advance in experimental entanglement demonstrations~\cite{Wilk2010,Isenhower2010,Zhang2010,Maller2015,Jau2015,Zeng2017,Levine2018,Picken2018,Jo2019} and high-fidelity quantum control~\cite{Levine2019,Graham2019,Madjarov2020}, neutral atoms emerged as a promising platform for large-scale quantum computing~\cite{PhysRevLett.85.2208,Lukin2001,Saffman2010,Saffman2016,Weiss2017,Adams2019,Wu2021,Morgado2021,Shi2021qst}. However, it is unclear whether it is possible to create controllable HE with neutral atoms. Until now, most entanglement experiments with individually trapped atoms~\cite{Wilk2010,Isenhower2010,Zhang2010,Maller2015,Jau2015,Zeng2017,Picken2018,Jo2019,Levine2019,Graham2019} focused on entanglement between hyperfine-Zeeman substates though the one in Ref.~\cite{Madjarov2020} studied entanglement between (electronic) Rydberg and clock states, and most theoretical studies on neutral-atom entanglement were also about hyperfine-Zeeman substates~\cite{Shi2021qst}. Rydberg interactions can lead to entanglement between the internal states and the motional states~\cite{Cozzini2006}, but the entanglement is difficult to control and more often appears as noise~\cite{Robicheaux2021}.

Here, we study HE operations with neutral atoms, namely, entanglement in both the electronic degree and the nuclear spin degree of neutral atoms. In particular, we consider neutral atoms whose outermost shell has two valence electrons, e.g., some alkaline-earth-metal or lanthanide atoms, which we call alkaline-earth-like~(AEL) atoms. Most AEL atoms have two stable electronic states, the ground $^1S_0$ state and the long-lived $^3P_0$ clock state. These two states can be the two states $|0(1)\rangle_{\text{e}}$ of a quantum bit~(qubit). Meanwhile, if the AEL atom possesses a nonzero nuclear spin, one can choose two nuclear spin states $|0(1)\rangle_{\text{n}}$ to define another qubit, where the subscript e and n denote the electronic and nuclear spin degrees of freedom, respectively. Then, the state of one atom is
\begin{eqnarray}
  (\cos\theta|0\rangle_{\text{e}}+e^{i\theta'}\sin\theta |1\rangle_{\text{e}} )\otimes(\cos\phi|0\rangle_{\text{n}}+e^{i\phi'}\sin\phi |1\rangle_{\text{n}} ), \nonumber\\
  \label{initialstate}
\end{eqnarray}
where $\theta,\theta',\phi$, and $\phi'$ are real variables, and the symbol $\otimes$ is used because the electronic and nuclear spin states are decoupled in the ground and clock states of the AEL atom we study. For the simplest case of two atoms, if quantum operations exist to entangle both the electronic and nuclear degrees of freedom, HE emerges. In this article, we show that it is possible to realize the controlled-Z~(C$_{\text{Z}}$) operation in both degrees of freedom. For two qubits where their initial state is $( |00\rangle_{\text{e}}+|01\rangle_{\text{e}}+|10\rangle_{\text{e}}+|11\rangle_{\text{e}})\otimes ( |00\rangle_{\text{n}}+|01\rangle_{\text{n}}+|10\rangle_{\text{n}}+|11\rangle_{\text{n}})/4$ the C$_{\text{Z}}\otimes$C$_{\text{Z}}$ operation in the electronic and nuclear states lead to $( |00\rangle_{\text{e}}+|01\rangle_{\text{e}}+|10\rangle_{\text{e}}-|11\rangle_{\text{e}})\otimes ( |00\rangle_{\text{n}}+|01\rangle_{\text{n}}+|10\rangle_{\text{n}}-|11\rangle_{\text{n}})/4$, which is an HE state where the electronic qubits in the two atoms are maximally entangled, and meanwhile the nuclear qubits are maximally entangled as well.

A recent work~\cite{Shi2021} showed that it is possible to use Rydberg blockade to entangle nuclear spin Zeeman substates in the ground level of divalent neutral atoms, but the theory therein can not lead to HE. An outstanding challenge to realize electron-nuclear spin HE in neutral atoms is that the electronic and nuclear spin states are decoupled in the ground and clock states~(which is true for most cases except some exceptions such as $^{165}$Ho~\cite{Saffman2008}), but when the states are Rydberg excited, the electronic and nuclear spin states are coupled.

The remainder of this article is organized as follows. In Sec.~\ref{section02}, we study the C$_{\text{Z}}$ operation with electronic qubits defined by the ground $^1S_0$ state and the stable $^3P_0$ clock state. In Sec.~\ref{section03}, we study the C$_{\text{Z}}$ operation in the nuclear spin states. In Sec.~\ref{section04}, we analyze $^{87}$Sr and $^{171}$Yb about the experimental prospects to realize the key steps in our theories. Section~\ref{section05} studies realization of single-qubit operations, Sec.~\ref{section06} discusses entanglement within one atom and between electronic states in one atom and nuclear spin states in another, and Sec.~\ref{section07} gives a brief summary.

\section{C$_{\text{Z}}$ gates with electronic qubits}\label{section02}
HE in this article is created by sequentially entangling the electronic states and the nuclear spin states. In this section, we study the method to entangle the electronic states without changing the nuclear spin states. 

\subsection{Challenges in realizing electronic C$_{\text{Z}}$ operations}\label{section02A}
It looks difficult to realize an entangling gate in the electronic qubits when there are also nuclear spin qubits in the atoms, i.e., when each electronic state is a superposition of different nuclear spin states. The issue stems from that both nuclear spin qubit states shall be excited for each step of the electronic state excitation. The Rydberg excitation of AEL atoms was experimentally achieved in~\cite{Madjarov2020} without involving nuclear spins for nuclear-spin-free $^{88}$Sr was used in~\cite{Madjarov2020}. 

The issue is understood as follows. For an AEL atom with a nonzero nuclear spin, because we not only use electronic states, namely, the ground $^1S_0$ state and a metastable clock~(the lowest excited) $^3P_0$ state to define qubit states $|0(1)\rangle_{\text{e}}$, we also use nuclear spin states~(with nuclear spin projections, e.g., $I$ and $I-1$, along the external magnetic field) $|0(1)\rangle_{\text{n}}$ to define another qubit, the general state for either the control or the target atom is
\begin{eqnarray}
 (\cos\theta|0\rangle_{\text{e}}+\sin\theta |1\rangle_{\text{e}} )\otimes(\cos\phi|0\rangle_{\text{n}}+\sin\phi |1\rangle_{\text{n}} ),\label{initialstate1}
\end{eqnarray}
where we ignore a relative phase between $|0\rangle$ and $|1\rangle$ which appeared in Eq.~(\ref{initialstate}). The state is shown by a product of the electronic and nuclear spin states in Eq.~(\ref{initialstate}) because the nuclear spin is decoupled from the electrons in the ground state; for the clock state there is a tiny mixing of the singlet states~\cite{Boyd2007}, and the nuclear spin is decoupled from the electrons for a first approximation. For frequently studied AEL atoms like ytterbium~\cite{Yamamoto2016,Saskin2018,wilson2019trapped} and strontium~\cite{Madjarov2020,Cooper2018,Covey2019,Norcia2018,Teixeira2020}, the electronic qubit states $|0\rangle_{\text{e}}$ and $|1\rangle_{\text{e}}$ are well separated by hundreds of THz. For the nuclear spin qubit states, the two states $|0(1)\rangle_{\text{n}}$ are separated by the Zeeman splitting $g_I\mu_n B$, where $g_I$ is the nuclear g factor, $\mu_n$ is the nuclear magnetic moment, and $B$ is the magnetic field. The value of $\mu_n$ is on the order of the nuclear magneton $\mu_N$ for both $^{87}$Sr~\cite{Sansonetti2010} and $^{171}$Yb~\cite{Porsev2004}, so that for a magnetic field $B$ on the order of Gauss~($1$~G$=10^{-4}$~T) as in experiments~\cite{Wilk2010,Zhang2010,Maller2015,Jau2015,Zeng2017,Levine2018,Picken2018,Levine2019,Graham2019}, the splitting between $|0\rangle_{\text{n}}$ and $|1\rangle_{\text{n}}$ is on the order of kHz which is useful to distinguish the two nuclear spin qubit states. When we say that the electronic state, e.g., $|0\rangle_{\text{e}}$, is excited to a Rydberg state, what we actually mean is that the state
\begin{eqnarray}
|0\rangle_{\text{e}}\otimes(\cos\phi|0\rangle_{\text{n}}+\sin\phi |1\rangle_{\text{n}} )\label{state01}
\end{eqnarray}
is excited to Rydberg states. Unfortunately, the state components $|0\rangle_{\text{e}}\otimes|0\rangle_{\text{n}}$ and $|0\rangle_{\text{e}}\otimes|1\rangle_{\text{n}}$ are related with two different nuclear spin projections. Both of them respond to the laser excitation in the form of electric dipole coupling, leading to Rydberg excitation. However, because hyperfine interaction will occur for Rydberg states and $|0\rangle_{\text{n}}$ and $|1\rangle_{\text{n}}$ have different nuclear spin projections, the two components $|0\rangle_{\text{e}}\otimes|0\rangle_{\text{n}}$ and $|0\rangle_{\text{e}}\otimes|1\rangle_{\text{n}}$ will be excited to Rydberg states with different $m_F$. There is in general strong singlet-triplet coupling for the $s-$orbital Rydberg states~\cite{Lurio1962,Lehec2018,Ding2018}, so that $|^1S_0,F=I\rangle$ is mixed with $|^3S_1,F=I\rangle$. Thus, whether we excite the qubit states to $|^3S_1,F=I\pm1\rangle$ or the mixed state of $|^1S_0,F=I\rangle$ and $|^3S_1,F=I\rangle$, the Zeeman splitting between two Rydberg states with $m_F$ differing by 1 is on the order of megahertz for a magnetic field on the order of $B\sim1$~G. This basically means that the Rydberg excitation for $|0\rangle_{\text{e}}\otimes|0\rangle_{\text{n}}$ and that for $|0\rangle_{\text{e}}\otimes|1\rangle_{\text{n}}$ can not be resonant simultaneously. If one is resonant, the other will be off-resonant with a MHz-scale detuning. But unfortunately, the Rabi frequency for the Rydberg excitation can not be very large, and values of several megahertz are already very large~\cite{Wilk2010,Isenhower2010,Zhang2010,Maller2015,Jau2015,Zeng2017,Levine2018,Picken2018,Levine2019,Graham2019,Madjarov2020}. Meanwhile, larger Rydberg Rabi frequencies are desirable for faster quantum control so as to suppress decoherence in the atomic systems. So, when the excitation for one of the two states $|0\rangle_{\text{e}}\otimes|0\rangle_{\text{n}}$ and $|0\rangle_{\text{e}}\otimes|1\rangle_{\text{n}}$ is resonant, the other will be excited with a detuning of a similar magnitude to the Rydberg Rabi frequency in terms of a generalized Rabi oscillation~\cite{Shi2017,Shi2018Accuv1,Shi2018prapp2,Levine2019,Shi2021}. Thus, it is impossible to use usual methods to excite one electronic qubit state to the Rydberg state without disturbing the other qubit state.

\subsection{Theory for Rydberg excitation of both nuclear spin qubit states}\label{section02B}
We study methods to fully excite both $|0\rangle_{\text{e}}\otimes|0\rangle_{\text{n}}$ and $|0\rangle_{\text{e}}\otimes|1\rangle_{\text{n}}$ to Rydberg states. Because $|0\rangle_{\text{e}}\otimes|0\rangle_{\text{n}}$ and $|0\rangle_{\text{e}}\otimes|1\rangle_{\text{n}}$ have different nuclear spin projections along the quantization axis, they are excited to Rydberg states of a common principal quantum number but different $m_F$. We label these two states by $|r_0\rangle$ and $|r_1\rangle$ corresponding to the two transitions
\begin{eqnarray}
  |0\rangle_{\text{e}}\otimes|0\rangle_{\text{n}}&\xrightarrow{\Omega_0}& |r_0\rangle,\nonumber\\
  |0\rangle_{\text{e}}\otimes|1\rangle_{\text{n}}&\xrightarrow{\Omega_1}& |r_1\rangle. \label{twotransitions}
\end{eqnarray}
where $\Omega_{0(1)}$ is the Rabi frequency for the corresponding transition. For highly excited states, the electronic and nuclear spin states are coupled so that we use $|r_{0(1)}\rangle$ to denote the Rydberg states. For a magnetic field $B$ on the order of Gauss, the splitting between $|0\rangle_{\text{e}}\otimes|0\rangle_{\text{n}}$ and $|0\rangle_{\text{e}}\otimes|1\rangle_{\text{n}}$ is on the order of kilohertz. As shown later, we consider that $|r_0\rangle$ and $|r_1\rangle$  are either $^3S_1$ Rydberg states, or superpositions of $^3S_1$ and $^1S_0$ Rydberg states due to hyperfine interactions. In this case, the Zeeman splitting between two Rydberg states with $m_F$ differing by 1 is on the order of megahertz for $B\sim1$~G. As a consequence, in the rotating frame, if one of the two transitions in Eq.~(\ref{twotransitions}) is resonant, the other will be off resonant with a detuning $\Delta$ on the order of megahertz. For an off-resonant transition between a ground state and a Rydberg state, e.g., the second in Eq.~(\ref{twotransitions}), a direct analysis based on the unitary dynamics shows that starting from the ground state, the population in the Rydberg state can not exceed $\Omega_1^2/(\Omega_1^2+\Delta^2)$~\cite{Shi2017,Shi2018Accuv1,Shi2018prapp2,Shi2021}. One can use a large magnetic field to increase $\Delta$ so as to suppress the unwanted transition. However, to have errors smaller than $10^{-4}$, we need $\Delta/\Omega_1$ on the order of $100$, which requires a large magnetic $B$ field. Nevertheless, large magnetic fields are related with larger spatial fluctuation that leads to strong dephasing~\cite{Saffman2005,Saffman2011}. For Rydberg entanglement experiments with ground-state atoms, the magnitudes of $B$ fields were smaller than $9$~G~\cite{Wilk2010,Zhang2010,Maller2015,Jau2015,Zeng2017,Levine2018,Picken2018,Levine2019,Graham2019} although Ref.~\cite{Isenhower2010} used a relatively large field of 11.5~G. Below, we show that there are ways to conquer the issue of Rydberg excitation with weak magnetic fields; in particular, we consider $|\Delta/\Omega_{0(1)}|\leq10$.

\begin{figure}
\includegraphics[width=2.5in]
{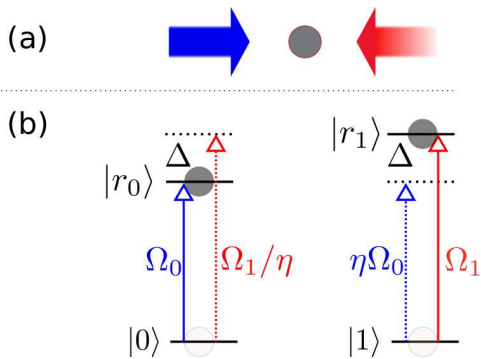}
\caption{ (a) An atom is excited by two sets of laser fields, one resonant with the transition from the qubit state $|0\rangle$ to the Rydberg state $|r_0\rangle$, and the other is resonant with the transition from $|1\rangle$ to $|r_1\rangle$. For the study in Sec.~\ref{subsection02B01}, the two states $|0\rangle$ and $|1\rangle$ refer to $|0\rangle_{\text{e}}\otimes|0\rangle_{\text{n}}$ and $|0\rangle_{\text{e}}\otimes|1\rangle_{\text{n}}$, respectively, but the theory in this figure applies to similar level diagrams in general. (b) Each set of fields not only couple the resonant transition, but also couple the off-resonant transition. Because of the selection rules, there can be a coefficient $\eta$ risen from the difference in the dipole couplings in the two transitions. As a result, the laser field resonant for the transition $|0\rangle\leftrightarrow|r_0\rangle$ with a Rabi frequency $\Omega_0$ will induce a detuned transition $|1\rangle\xleftrightarrow{\eta\Omega_0 e^{-it\Delta}}|r_1\rangle$. Meanwhile, the laser field resonant for the transition $|1\rangle\leftrightarrow|r_1\rangle$ with a Rabi frequency $\Omega_1$ will induce a detuned transition $|0\rangle\xleftrightarrow{\Omega_1 e^{it\Delta}/\eta}|r_0\rangle$. Here, $\Delta$ is the frequency separation between the two transitions from $|0(1)\rangle$ to the Rydberg states $|r_{0(1)}\rangle$.   \label{figure1} }
\end{figure}

\subsubsection{One-step Rydberg excitation with two lasers, one resonant and the other off-resonant }\label{subsection02B01}
It has been shown that when more than one Rabi frequencies control the Rydberg excitation from the ground state, resonance can arise from off resonance; see, e.g., Ref.~\cite{ShiJPB2020}. The method in Ref.~\cite{ShiJPB2020} was derived from a much earlier work~\cite{Goreslavsky1980}, where two symmetrically detuned excitation fields can lead to a resonant state excitation. To solve the problem of simultaneously exciting two nuclear spin states $|0\rangle_{\text{n}}$ and $|1\rangle_{\text{n}}$ to Rydberg states, we extend the method in Refs.~\cite{ShiJPB2020,Goreslavsky1980} and consider the scheme in Fig.~\ref{figure1}. Two laser fields are used to excite the states to Rydberg states, one resonant with the transition $|0\rangle_{\text{e}}\otimes|0\rangle_{\text{n}}\leftrightarrow|r_0\rangle$ by a Rabi frequency $\Omega_0$, while the other is resonant with the transition $ |0\rangle_{\text{e}}\otimes|1\rangle_{\text{n}}\leftrightarrow{}|r_1\rangle$ by a Rabi frequency $\Omega_1$. The two laser fields have the same polarization but have a frequency difference characterized by a detuning $\Delta$ in one of two transitions for each state. Then, because the values of $m_F$  of the two states $|0\rangle_{\text{e}}\otimes|0\rangle_{\text{n}}$ and $ |0\rangle_{\text{e}}\otimes|1\rangle_{\text{n}}$ differ by $1$, the ratio $\eta$ between the Rabi frequencies for the two transitions $ |0\rangle_{\text{e}}\otimes|0(1)\rangle_{\text{n}}\leftrightarrow{}|r_{0(1)}\rangle$ for each set of laser fields is usually not equal to 1. The Hamiltonians for the two transitions are respectively given by 
\begin{eqnarray}
  \hat{H}_0(t)&=& \frac{1}{2}   \left(
\begin{array}{cc}
    0& \Omega_0 + \Omega_1 e^{-i\Delta t}/\eta\\
\Omega_0 + \Omega_1 e^{i\Delta t}/\eta&0\end{array}
  \right) \label{RydbergPumptheory01}
\end{eqnarray}
with the basis $\{|r_{0}\rangle,~|0\rangle_{\text{e}}\otimes|0\rangle_{\text{n}}\}$, and 
\begin{eqnarray}
  \hat{H}_1(t)&=& \frac{1}{2}   \left(
\begin{array}{cc}
    0& \eta \Omega_0 e^{i\Delta t}+ \Omega_1\\
  \eta   \Omega_0 e^{-i\Delta t}+ \Omega_1 &0\end{array}
  \right) \label{RydbergPumptheory02}
\end{eqnarray}
with the basis $\{|r_{1}\rangle,~|0\rangle_{\text{e}}\otimes|1\rangle_{\text{n}}\}$. Here, $\eta$ is determined by the electric dipole selection rules and is assumed to be real for brevity.

An intuitive look at Eqs.~(\ref{RydbergPumptheory01}) and~(\ref{RydbergPumptheory02}) gives us an impression that it seems impossible to simultaneously excite both $|0\rangle_{\text{e}}\otimes|0\rangle_{\text{n}}$ and $|0\rangle_{\text{e}}\otimes|1\rangle_{\text{n}}$ to Rydberg states. However, if the Rabi frequency $\Omega_{0(1)}$ is time dependent in the form of 
\begin{eqnarray}
   \Omega_{0(1)} &=& 2i\kappa_{0(1)}\sin(\delta t), \label{RydbergPumptheory03}
\end{eqnarray}
Rydberg excitation can still proceed in the limit $\delta\ll\Delta$, where $\kappa_{0(1)}$ is a positive frequency for brevity. A similar analysis was shown in Ref.~\cite{ShiJPB2020} where only $\kappa_{0(1)}$ is present for the excitation of the state $|0(1)\rangle$. 

\begin{figure}
\includegraphics[width=3in]
{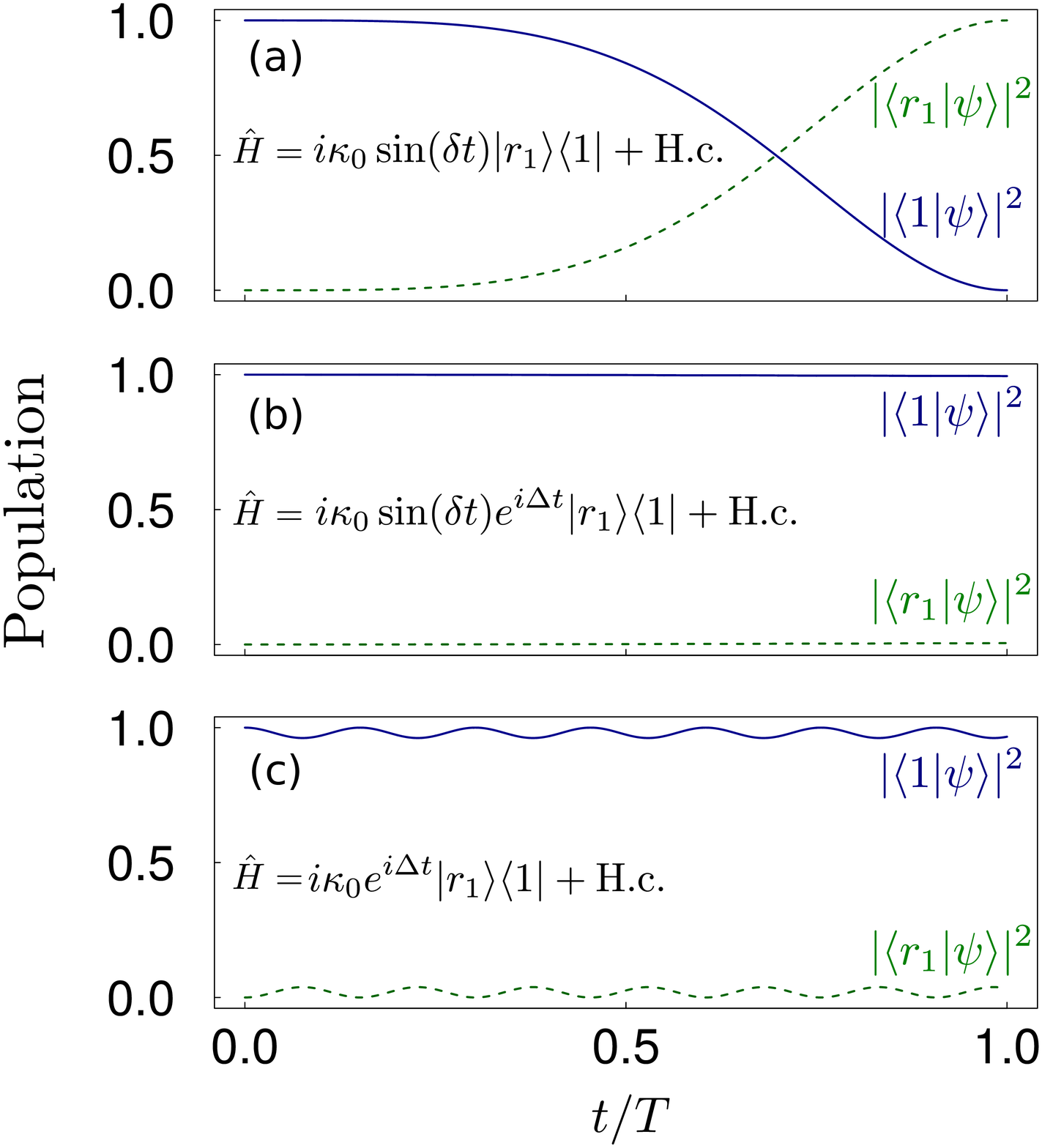}
\caption{ Population dynamics for $|1\rangle$ and $|r_1\rangle$ in a two-level system with three different Hamiltonians.  (a) With a Hamiltonian $i\kappa_{0}\sin(\delta t) |r_1\rangle\langle1|+\text{H.c.}$, a complete population inversion occurs with $T=$acos$[1-\pi\delta/(2\kappa_0)]/\delta$. (b) If the laser fields are largely detuned with $\Delta$, the population in $|r_1\rangle$ is tiny during the pulse. (c) If rectangular pulses are used, i.e., we use the Hamiltonian $i\kappa_{0}e^{i\Delta t} |r_1\rangle\langle1|+\text{H.c.}$, the population error or leakage in the Rydberg state is $0.034$ at the end of the pulse. The condition $(\Delta,~\delta)/(2\kappa_0) = (5,~0.1)$ is used here.   \label{comparison} }
\end{figure}

To understand the Rydberg excitation by the fields shown in Fig.~\ref{figure1} with condition~(\ref{RydbergPumptheory03}), we start from the case when only the excitation with the Rabi frequency $\Omega_{0}$ is present, then Eq.~(\ref{RydbergPumptheory01}) becomes
\begin{eqnarray}
  \hat{H}_0'&=& \frac{1}{2}   \left(
\begin{array}{cc}
    0& 2i\kappa_{0}\sin(\delta t) \\
-2i\kappa_{0}\sin(\delta t)&0\end{array}
  \right) .\label{H0anotherform}
\end{eqnarray}
By evaluating the time ordering operator, one can find that starting from the ground state $|\psi(t=0)\rangle=|0\rangle_{\text{e}}\otimes|0\rangle_{\text{n}}$, the amplitude in $|r_0\rangle$ is
\begin{eqnarray}
 \langle r_0|\psi(t)\rangle = \sin\frac{\kappa_0[1-\cos(\delta t)]}{\delta},
\end{eqnarray}
while $ \langle \psi(t=0)|\psi(t)\rangle = \cos\frac{\kappa_0[1-\cos(\delta t)]}{\delta}$, which means that at the moment $t=T$ with $T$ given by 
\begin{eqnarray}
\frac{\kappa_0[1-\cos(\delta T)]}{\delta}=\frac{\pi}{2},\label{sinpumping}
\end{eqnarray}
the Rydberg state $|r_0\rangle$ is fully populated. A numerical simulation about this phenomenon is shown in Fig.~\ref{comparison}(a) which shows that the final state is indeed $|r_0\rangle$ if we identify the basis $\{|r_1\rangle,|1\rangle\}$ in Fig.~\ref{comparison}(a) as the basis of Eq.~(\ref{H0anotherform}). To avoid interrupting the state $|0\rangle_{\text{e}}\otimes|1\rangle_{\text{n}}$ by the fields for realizing the transition $|0\rangle_{\text{e}}\otimes|0\rangle_{\text{n}}\leftrightarrow|r_0\rangle$ as discussed above, one can set $\delta\ll \Delta$. The reason is as follows. When the fields act on the transition from $|0\rangle_{\text{e}}\otimes|1\rangle_{\text{n}}$ to the Rydberg state, there is not only a coefficient $\eta$ as shown in Eq.~(\ref{RydbergPumptheory02}), but also a large detuning $\Delta$. Then, the Hamiltonian for $|0\rangle_{\text{e}}\otimes|1\rangle_{\text{n}}$ is 
\begin{eqnarray}
  \hat{H}_1'&=& \hat{H}_{1+}'+\hat{H}_{1-}',\nonumber\\
  \hat{H}_{1+}'&=& \frac{\eta}{2}   \left(
\begin{array}{cc}
    0& \kappa_{0}e^{i(\delta+\Delta) t} \\
\kappa_{0}e^{-i(\delta+\Delta) t} &0\end{array}
  \right) ,\label{positivepart2} \\
  \hat{H}_{1-}'&=& \frac{\eta}{2}   \left(
\begin{array}{cc}
    0& -\kappa_{0} e^{i(\Delta-\delta) t} \\
-\kappa_{0}e^{-i(\Delta-\delta) t}&0\end{array}
  \right) .\label{negativepart2}
\end{eqnarray}
The two highly off-resonant transitions in Eqs.~(\ref{positivepart2}) and~(\ref{negativepart2}) are with slightly different detunings, but for the condition $\delta\ll \Delta$, it is like that both Eqs.~(\ref{positivepart2}) and~(\ref{negativepart2}) have the same detuning $\Delta$, which simply means that $\hat{H}_{1+}+\hat{H}_{1-}=0$. So, the state $|0\rangle_{\text{e}}\otimes|1\rangle_{\text{n}}$ seems to feel no fields, and, hence, not excited. 

The key point in the above discussion lies in that without using large $\Delta/\kappa_0$ we can suppress the Rydberg excitation of $|0\rangle_{\text{e}}\otimes|1\rangle_{\text{n}}$. To show this, we consider $\eta=1$ for brevity and $(\Delta,~\delta)/(2\kappa_0) = (5,~0.1)$ as an example, and compare the off-resonant excitation by the Hamiltonian $\hat{H}_1'$~[which is equal to the sum of Eq.~(\ref{positivepart2}) and~(\ref{negativepart2})] and the excitation by the following Hamiltonian
\begin{eqnarray}
  \hat{H}_1''&=& \frac{1}{2}   \left(
\begin{array}{cc}
    0& 2i\kappa_{0}e^{i\Delta t} \\
-2i\kappa_{0}e^{-i\Delta t}&0\end{array}
  \right) .\label{H0anotherform03}
\end{eqnarray}
As numerically shown in Fig.~\ref{comparison}(b), the leakage to the Rydberg state is negligible with Hamiltonian $i\kappa_{0}\sin(\delta t)e^{i\Delta t} |r_1\rangle\langle1|+\text{H.c.}$, but when Eq.~(\ref{H0anotherform03}) is used, the leakage can be as large as $3\%$ shown in Fig.~\ref{comparison}(c). More than the error in the population, the final phase of $\langle1|\psi\rangle$ in Fig.~\ref{comparison}(c) is as large as $0.13\pi$, while that in Fig.~\ref{comparison}(b) is only $0.025\pi$. The phase error is quite detrimental concerning the realization of a controlled-phase gate such as C$_{\text{Z}}$~\cite{Shi2020prapplied}. These data show that the application of a field that is largely detuned and slowly varying, which has a Hamiltonian in the form of Eq.~(\ref{H0anotherform}) for the targeted transition, is advantageous for suppressing the detrimental influence on the other qubit state that is off-resonantly coupled.

\begin{figure}
\includegraphics[width=3in]
{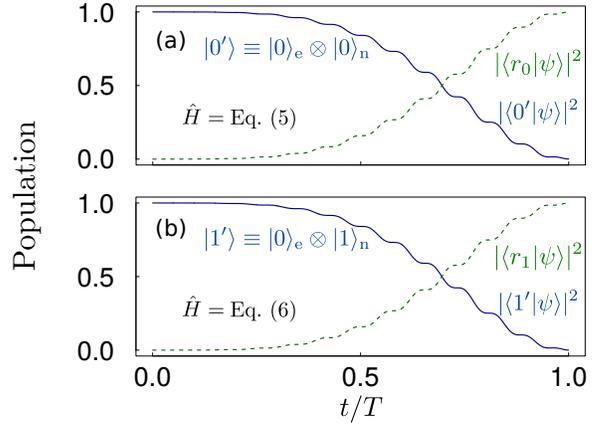}
\caption{ Population evolution in the ground and Rydberg states, shown by solid and dashed curves, respectively. (a) and (b) show results with the Hamiltonians in Eq.~(\ref{RydbergPumptheory01}) and in Eq.~(\ref{RydbergPumptheory02}), respectively. The pulse duration is $T=$acos$[1-\pi\delta/(2\kappa_0)]/\delta$, where $\kappa_0=\kappa_1$ and $(\Delta,~\delta)/(2\kappa_0) = (10,~0.1)$. The final population in $|r_{0(1)}\rangle$ is $0.99985$ in both (a) and (b). The phase arg$\langle r_{0(1)}|\psi\rangle$ is $\pm0.0028\pi$ in (a)~[(b)] at the end of the pulse, while the desired phase is 0.   \label{comparison2} }
\end{figure}

Finally, we study Rydberg excitations of $|0\rangle_{\text{e}}\otimes|0\rangle_{\text{n}}$ and $|0\rangle_{\text{e}}\otimes|1\rangle_{\text{n}}$ with the two laser fields shown in Fig.~\ref{figure1}. Because each of the two sets of fields are resonant with one transition only, the relevant Hamiltonians are given in Eqs.~(\ref{RydbergPumptheory01}) and~(\ref{RydbergPumptheory02}). For brevity, we consider the case $\eta=1$ because a value different from 1 only has a marginal influence on the error shown in, e.g., Fig.~\ref{comparison}(b). The time dependence for $\Omega_{0(1)}$ is specified in Eq.~(\ref{RydbergPumptheory03}). To have a high-fidelity state excitation, we use the condition $(\Delta,~\delta)/(2\kappa_0) = (10,~0.1)$ to simulate the population evolution by Eq.~(\ref{RydbergPumptheory01}) and Eq.~(\ref{RydbergPumptheory02}), with results shown in Fig.~\ref{comparison2}(a) and~\ref{comparison2}(b), respectively. The population error is smaller than $1.5\times10^{-5}$ in both cases, and the phase error~(the correct phase should be $0$) is smaller than $0.003\pi$. So, it is a useful way to use the scheme shown in Fig.~\ref{figure1} for Rydberg excitation of the electronic qubits when there is a nonzero nuclear spin in the atom. If the initial state is $|r_{0(1)}\rangle$ for the Hamiltonian in Eq.~(\ref{RydbergPumptheory01})[~(\ref{RydbergPumptheory02})], a similar pulse can deexcite them back to the ground states, where the final phase for the ground state is $\pi$, and numerical simulations give similar results about the errors in the population and phase.


\begin{figure*}[ht]
\includegraphics[width=5.0in]
{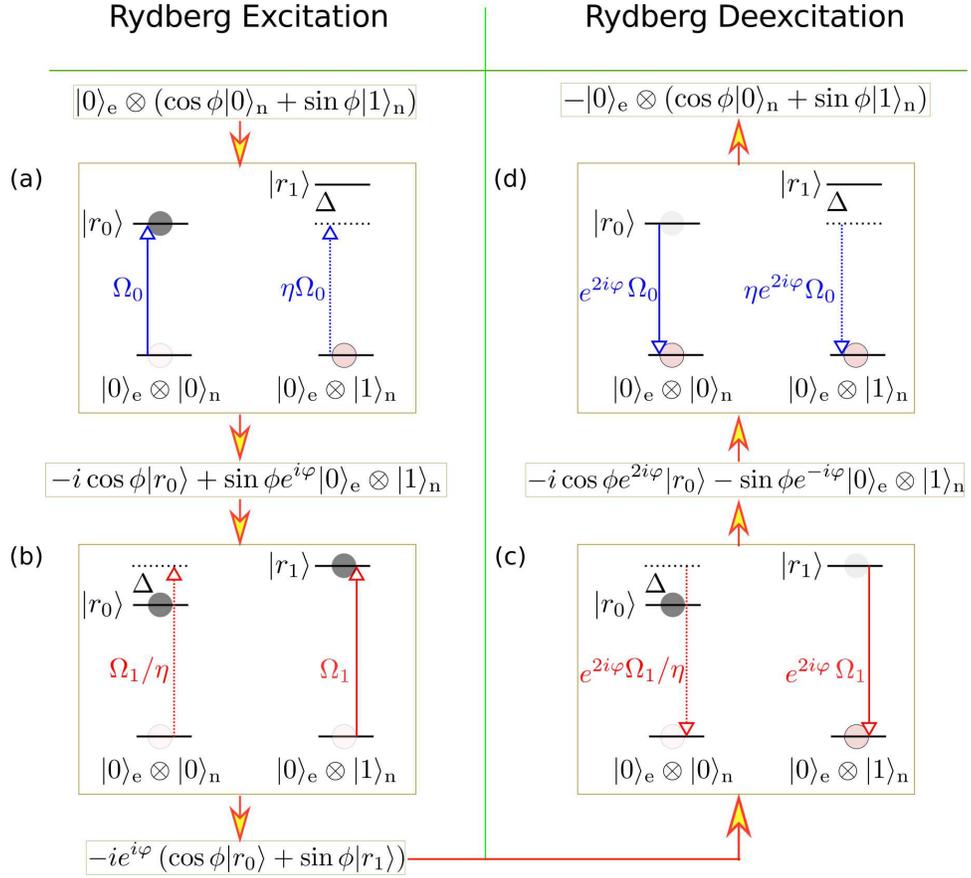}
\caption{A two-step Rydberg excitation scheme, and a two-step deexcitation scheme. Symbols beside the arrows denote Rabi frequencies, and the directions of arrows denote the direction of population transfer. (a) An atom is excited by laser fields that are resonant with the transition from $|0\rangle_{\text{e}}\otimes|0\rangle_{\text{n}}$ to the Rydberg state $|r_0\rangle$. It induces an off-resonant excitation from $|0\rangle_{\text{e}}\otimes|1\rangle_{\text{n}}$ to $|r_1\rangle$. With $\overline{\Omega}\equiv\sqrt{\eta^2\Omega_{0}^2+\Delta^2}$ equal to $2N$ times $\Omega_{0}$, where $N$ is an integer, the state $|0\rangle_{\text{e}}\otimes|1\rangle_{\text{n}}$ returns to itself with a phase twist. (b) When the laser fields are set to be resonant with the transition from $|0\rangle_{\text{e}}\otimes|1\rangle_{\text{n}}$ to the Rydberg state $|r_1\rangle$, and the Rabi frequencies are chosen so that the generalized Rabi frequency for the transition between $|0\rangle_{\text{e}}\otimes|0\rangle_{\text{n}}$ and $|r_0\rangle$ is $2N$ times $\Omega_1$, the state $|r_0\rangle$ returns to itself with a phase change. The two pulses in (a) and (b) induce a full Rydberg excitation. Here, (c) shows the state evolution with the same pulse as in (b) except a phase $2\varphi$ in the Rabi frequency, which deexcites the state $|r_1\rangle$ back to the ground state. Likewise, (d) shows the Rydberg deexcitation for $|r_0\rangle$, which uses laser fields as used in (a) except a phase $2\varphi$ in the Rabi frequency.  \label{figure4} }
\end{figure*}

\subsubsection{A two-step Rydberg excitation by the method of Ref.~\cite{Shi2021} }\label{subsection02B02}
The method in Sec.~\ref{subsection02B01} requires $\Delta/\Omega_{0(1)}$ as large as, for example, $10$, as in Fig.~\ref{comparison2}. If large Rydberg Rabi frequencies are realized, for example, as large as $2\pi\times7$~MHz~\cite{Madjarov2020}, it is challenging to use the theory in Sec.~\ref{subsection02B01} unless larger magnetic fields are employed so as to induce a large $\Delta$. Below, we describe another method that is compatible with small $\Delta/\Omega_{0(1)}$ but can also lead to high-fidelity Rydberg excitation of both nuclear spin states. The method is derived from the theory shown in Ref.~\cite{Shi2021} which studies selective Rydberg excitation of a nuclear spin state in the ground-state manifold. 

The theory is shown in Fig.~\ref{figure4}, which can be understood as a repeat of the Rydberg excitation scheme in Ref.~\cite{Shi2021} with specific detunings. As an example, we consider exciting the state $|0\rangle_{\text{e}}\otimes|1\rangle_{\text{n}}$ to Rydberg state with a constant Rabi frequency $\Omega_0$. The fields can induce a Rydberg Rabi frequency $\eta\Omega_0$ for the transition $|0\rangle_{\text{e}}\otimes|1\rangle_{\text{n}}\leftrightarrow|r_1\rangle$ shown in Fig.~\ref{figure4}(a), where $\eta$ is from the angular momentum coupling rules. $\eta$ can be a complex variable, but whether being real or complex will not have nontrivial effects in a complete cycle of detuned Rabi oscillation as we consider in this work. So, we assume real $\eta$ for brevity. When $\pi$ polarized fields are used with $^{171}$Yb atoms, $|\eta|$ is 1 because the two nuclear spin states are symmetrical to each other except a sign difference in the coupling matrices~(see, e.g, the appendixes of Ref.~\cite{Shi2021}). For AEL qubits whose nuclear pin $I$ is larger than $1/2$, the value of $|\eta|$ is not 1 in general but it does spoil our theory. For brevity, we consider $\eta=1$. The Hamiltonians for the two states $|0\rangle_{\text{e}}\otimes|0\rangle_{\text{n}}$ and $|0\rangle_{\text{e}}\otimes|1\rangle_{\text{n}}$ are given by Eqs.~(\ref{RydbergPumptheory01}) and~(\ref{RydbergPumptheory02}), respectively, with $\Omega_1=0$. To analytically derive the time dynamics, we use appropriate rotating frames with operators
\begin{eqnarray}
 \hat{R}&=&\sum_{j}E_j|j\rangle\langle j|,\nonumber\\
 \hat{R}_0&=&\hat{R}+\Delta|r_0\rangle\langle r_0|,\nonumber\\
 \hat{R}_1&=&\hat{R}-\Delta|r_1\rangle\langle r_1|,\label{rotatingOperators}
\end{eqnarray}
to transform the wavefunction $|\Psi\rangle$ in the Schr\"{o}dinger picture to wavefunctions $e^{it\hat{R}}|\Psi\rangle,~e^{it\hat{R}_0}|\Psi\rangle$, or $e^{it\hat{R}_1}|\Psi\rangle$ in rotating frames, where $|j\rangle$ and $E_j$ are the wavefunction and eigenenergy of the atomic state labeled by $j$. In particular, we study quantum gates in the frame rotating with $\hat{R}$, i.e., with wavefunctions $|\psi\rangle\equiv e^{it\hat{R}}|\Psi\rangle$, but will sometimes use the frames rotating with $\hat{R}_{0(1)}$ when analyzing the system dynamics. For the excitation in Fig.~\ref{figure4}(a), the off-resonant excitation on the state $|0\rangle_{\text{e}}\otimes|1\rangle_{\text{n}}$ is described by Eq~(\ref{RydbergPumptheory02}) which can be transformed to the frame defined by $\hat{R}_1$,
\begin{eqnarray}
  \hat{H}_1^{(\hat{R}_1)}&=& \frac{1}{2}   \left(
\begin{array}{cc}
   2 \Delta&  \Omega_0 \\
    \Omega_0  & 0  \end{array}
  \right) ,\label{RydbergPumptheory02rotate}
\end{eqnarray}
where $\Omega_0$ is constant instead of being time-dependent as in Sec.~\ref{subsection02B01}. The Hamiltonian above can be diagonalized as~\cite{Shi2017}
\begin{eqnarray}
  \hat{H}_{1}^{(\hat{R}_1)}&=& \sum_{\alpha=\pm} \epsilon_\alpha |v_\alpha\rangle\langle v_\alpha|,\nonumber
\end{eqnarray}
where
\begin{eqnarray}
  \epsilon_\pm &=& (\Delta \pm \sqrt{\Omega_0^2+\Delta^2})/2,\nonumber\\
  |v_\pm \rangle &=& \left(\frac{\Omega_0}{2} |0\rangle_{\text{e}}\otimes|1\rangle_{\text{n}} + \epsilon_\pm |r_1\rangle\right)/\mathcal{N}_\pm,\nonumber\\
  \mathcal{N}_\pm  &=& \sqrt{\Omega_0^2/4 + \epsilon_\pm^2},
\end{eqnarray}
from which we have
\begin{eqnarray}
|0\rangle_{\text{e}}\otimes|1\rangle_{\text{n}} &=&\frac{2}{\Omega_0}\frac{\epsilon_-\mathcal{N}_+ |v_+ \rangle - \epsilon_+\mathcal{N}_- |v_- \rangle  }{ \epsilon_-  - \epsilon_+ } ,\nonumber\\
  |r_1\rangle &=&\frac{\mathcal{N}_+ |v_+ \rangle - \mathcal{N}_- |v_- \rangle  }{ \epsilon_+  - \epsilon_- }. \nonumber
\end{eqnarray}
Starting from an initial state $|\psi(0)\rangle=|0\rangle_{\text{e}}\otimes|1\rangle_{\text{n}}$, the wavefunction becomes
\begin{eqnarray}
  |\psi(t)\rangle &=&\frac{2}{\Omega_0}\hat{\mathscr{S}}\frac{\epsilon_-\mathcal{N}_+e^{-it\epsilon_+} |v_+ \rangle - \epsilon_+\mathcal{N}_-e^{-it\epsilon_-} |v_- \rangle  }{ \epsilon_-  - \epsilon_+ } ,\label{v0pmvdwI}
\end{eqnarray}
where the frame transformation factor $\hat{\mathscr{S}}\equiv e^{it\hat{R}}e^{-it\hat{R}_1}$ can be expanded as $e^{it\Delta|r_1\rangle\langle r_1|}$. A $\pi$ pulse with duration $t_\pi=\pi/\Omega_0$ can complete the transition $|0\rangle_{\text{e}}\otimes|0\rangle_{\text{n}}\rightarrow-i|r_0\rangle$. To avoid exciting $|0\rangle_{\text{e}}\otimes|1\rangle_{\text{n}}$ to Rydberg states, it is desirable to realize a generalized Rabi frequency $\overline{\Omega} =\sqrt{\Omega_{0}^2+\Delta^2}$ which is $2N$ times $\Omega_{0}$, where $N$ is an integer. The reason that we call $\overline{\Omega}$ a generalized Rabi frequency~\cite{Shi2017} lies in that although full Rydberg excitation is not achieved, there is a state rotation from the ground state to a superposition of the ground state and the Rydberg state. According to Eq.~(\ref{v0pmvdwI}), with a pulse of duration $t_\pi$, the input state $|0\rangle_{\text{e}}\otimes|1\rangle_{\text{n}}$ undergoes $N$ detuned Rabi cycles,
\begin{eqnarray}
|0\rangle_{\text{e}}\otimes|1\rangle_{\text{n}} \rightarrow e^{i\varphi} |0\rangle_{\text{e}}\otimes|1\rangle_{\text{n}},\label{mappingof0}
\end{eqnarray}
where $\varphi\equiv-[N+\Delta/(2\Omega_{0}) ]\pi$; here the frame transformation factor $\hat{\mathscr{S}}$ does not take effect because there is no population in the Rydberg state. Thus, with a $\pi$ pulse which is resonant with the transition  
\begin{eqnarray}
|0\rangle_{\text{e}}\otimes|0\rangle_{\text{n}} \xrightarrow{\Omega_0} -i|r_0\rangle, \label{mappingof0-2}
\end{eqnarray}
the other nuclear spin state stays there, with only a phase accumulation as shown in Eq.~(\ref{mappingof0}). For a general qubit state initialized in
\begin{eqnarray}
|0\rangle_{\text{e}}\otimes(\cos\phi|0\rangle_{\text{n}}+\sin\phi|1\rangle_{\text{n}}),\label{theory2step00initial}
\end{eqnarray}
the first $\pi$ pulse changes it to~(in the frame $\hat{R}$)
\begin{eqnarray}
-i\cos\phi|r_0\rangle +\sin\phi e^{i\varphi} |0\rangle_{\text{e}}\otimes|1\rangle_{\text{n}},\label{theory2step01}
\end{eqnarray}
as schematically shown in Fig.~\ref{figure4}(a).

The state in Eq.~(\ref{theory2step01}) has not been fully excited to Rydberg state. In order to excite the component $|0\rangle_{\text{e}}\otimes|1\rangle_{\text{n}}$ to Rydberg state, we use the same strategy as used from Eq.~(\ref{RydbergPumptheory02rotate}) to Eq.~(\ref{theory2step01}), but with Rydberg laser fields resonant with 
\begin{eqnarray}
|0\rangle_{\text{e}}\otimes|1\rangle_{\text{n}} \xrightarrow{\Omega_1} -i|r_1\rangle, \label{theory2step2r1}
\end{eqnarray}
so the other state $|r_0\rangle$ will be off-resonantly excited to $|0\rangle_{\text{e}}\otimes|0\rangle_{\text{n}}$. In this case, Eq.~(\ref{RydbergPumptheory01}), with $\Omega_0=0$, is transformed to the frame rotating with $\hat{R}_0$, 
\begin{eqnarray}
  \hat{H}_0^{(\hat{R}_0)}&=& \frac{1}{2}   \left(
\begin{array}{cc}
   -2 \Delta&  \Omega_1 \\
    \Omega_1  & 0  \end{array}
  \right) ,
\end{eqnarray}
and the wavefunction $|r_0\rangle$ is transformed to $e^{it_\pi\hat{R}_0}e^{-it_\pi\hat{R}}|r_0\rangle=e^{it_\pi\Delta}|r_0\rangle$, where the magnitudes of the Rydberg laser fields are chosen so that $|\Omega_1|$ is equal to $|\Omega_0|$ in Eq.~(\ref{RydbergPumptheory02rotate}) and the pulse is applied in the time interval $t\in(t_\pi,~2t_\pi]$. By using the same theoretical scheme as above, a $\pi$ pulse will excite the state $|0\rangle_{\text{e}}\otimes|1\rangle_{\text{n}}$ to $|r_1\rangle$, but the state component $|r_0\rangle$ will experience $N$ detuned Rabi cycles, evolving as
\begin{eqnarray}
|r_0\rangle &\rightarrow&e^{2it_\pi\hat{R}}e^{-2it_\pi\hat{R}_0} [ e^{-i\varphi}(e^{it_\pi\Delta} |r_0\rangle)] \nonumber\\
  &=&e^{-i\varphi}e^{-it_\pi\Delta}  |r_0\rangle   ,\label{mappingof1}
\end{eqnarray}
where $\varphi$ is defined below Eq.~(\ref{mappingof0}). By using the fact $e^{2iN\pi}=1$, we have $e^{-i\varphi}e^{-it_\pi\Delta}=e^{i\varphi}$. As a consequence, the state in Eq.~(\ref{theory2step01}) evolves to~(in the frame $\hat{R}$)
\begin{eqnarray}
-ie^{i\varphi}\left(\cos\phi |r_0\rangle+\sin\phi  |r_1\rangle\right).\label{theory2step02}
\end{eqnarray}
The second step is schematically shown in Fig.~\ref{figure4}(b).

After fully exciting the state $|0\rangle_{\text{e}}\otimes(\cos\phi|0\rangle_{\text{n}}+\sin\phi|1\rangle_{\text{n}})$ to the Rydberg state in Eq.~(\ref{theory2step02}), the mechanism of Rydberg blockade can be implemented. For example, when a control qubit experiences the above two-step Rydberg excitation, its Rydberg population can block the Rydberg excitation of a nearby target qubit. After the quantum manipulation of the target qubit, the state of the control qubit shall be restored back to the ground state.

Similar to the Rydberg excitation, the deexcitation also needs two steps, shown in Figs.~\ref{figure4}(c) and~\ref{figure4}(d). However, if the Rabi frequencies are still those used in the excitation processes, the final state will have undesired phases. So, the third step is similar to the second step but with a Rabi frequency $\Omega_1e^{2i\varphi}$, where the phase in the Rabi frequency will not have any nontrivial effect for a full detuned Rabi cycle as shown in Eqs.~(\ref{RydbergPumptheory02rotate})-(\ref{mappingof0}), but will have a net effect in half of a {\it resonant} Rabi cycle. Then, the third pulse realizes
\begin{eqnarray}
 |r_0\rangle& \rightarrow&e^{i\varphi}|r_0\rangle, \nonumber\\
 |r_1\rangle& \rightarrow&-ie^{-2i\varphi}|0\rangle_{\text{e}}\otimes |1\rangle_{\text{n}}, \nonumber
\end{eqnarray}
so that the state in Eq.~(\ref{theory2step02}) evolves to~(in the frame $\hat{R}$)
\begin{eqnarray}
-i\cos\phi e^{2i\varphi}|r_0\rangle-\sin\phi e^{-i\varphi} |0\rangle_{\text{e}}\otimes |1\rangle_{\text{n}}.\label{theory2step03}
\end{eqnarray}
Likewise, the fourth step is similar to the first step, except that the Rabi frequency changes to $\Omega_0e^{2i\varphi}$. Then, it realizes
\begin{eqnarray}
 |r_0\rangle& \rightarrow&-ie^{-2i\varphi}|0\rangle_{\text{e}}\otimes |0\rangle_{\text{n}}, \nonumber\\
 |0\rangle_{\text{e}}\otimes |1\rangle_{\text{n}}& \rightarrow&e^{i\varphi}|0\rangle_{\text{e}}\otimes |1\rangle_{\text{n}}, \nonumber
\end{eqnarray}
so that the state in  Eq.~(\ref{theory2step03}) evolves to~(in the frame $\hat{R}$)
\begin{eqnarray}
-|0\rangle_{\text{e}}\otimes(\cos\phi|0\rangle_{\text{n}}+\sin\phi|1\rangle_{\text{n}}).\nonumber\label{theory2step04}
\end{eqnarray}

 \subsection{Application in the ground state and the clock state}
 In Secs.~\ref{subsection02B01} and~\ref{subsection02B02} the two methods can be applied to any electronic states of the form $^{2S+1}L_0$. For most AEL atoms with a nonzero nuclear spin, the ground $^1S_0$ state is long lived, and there is a metastable $^3P_0$ state that is long lived, too. The g factor for the ground state is simply the nuclear spin g factor, while that for the $^3P_0$ clock state is mainly from the nuclear spin but there is a $60\%$ enhancement from the singlet-triplet mixing due to the spin-orbit and hyperfine interactions for the case of $^{87}$Sr~\cite{Boyd2007}. For other elements the enhancement can differ due to different strengths in the spin-orbit and hyperfine interactions, but even so, the magnitude of the Zeeman shift in $^3P_0$ is of similar magnitude to that in the ground state. Because the nuclear g factor is three orders of magnitude smaller than that of the electron spin, the Zeeman shift in $^3P_0$ is still on the order of kHz. So, compared to the Zeeman shift in the Rydberg state, the two nuclear spin states in either the ground state $^1S_0$ or the clock state $^3P_0$ are nearly degenerate. As a result, the parameter $\Delta$ in Secs.~\ref{subsection02B01} and~\ref{subsection02B02} are mainly from the Zeeman shift of the Rydberg states, and, hence, the two theories are applicable to both electronic qubit states.

 One difference between exciting the two electronic qubit states to an $s$-orbital Rydberg state is that a two-photon excitation is required for the qubit state $^1S_0$, while only a one-photon excitation is required for the clock state $^3P_0$~\cite{Madjarov2020}.

 \subsection{C$_{\text{Z}}$ operation in the electronic states and its fidelity}\label{electronicCZ}
 The Rydberg excitation shown in Secs.~\ref{subsection02B01} and~\ref{subsection02B02} is the foundation of our quantum gates. To realize a C$_{\text{Z}}$ gate in the electronic qubits, one can use the following pulse sequence. First, apply the pulse sequence shown in Sec.~\ref{subsection02B01} or Sec.~\ref{subsection02B02} for the control atom. When the first theory is used, the laser configuration is shown in Fig.~\ref{figure1} and the pulse duration is shown in Eq.~(\ref{sinpumping}); when the second theory is used, the laser pulse for Rydberg excitation is shown in Fig.~\ref{figure4}(a) and~\ref{figure4}(b). Second, if the first theory is used, then apply the laser fields as in Fig.~\ref{figure1} for a duration~[see Eq.~(\ref{sinpumping}) and explanations around it]
\begin{eqnarray}
\frac{\kappa_0[1-\cos(T\delta)]}{\delta}=\pi,\label{pidurationtheory1}
\end{eqnarray}
so that a $\pi$ phase appears for the input state $|10\rangle$; if the second theory is used, then use the pulse sequence shown in Fig.~\ref{figure4} for the target atom which also induces a $\pi$ phase shift. For the input state $|10\rangle$ in the electronic state space, there will be a $\pi$ phase imprinted in it at the end of the four-pulse sequence shown in Fig.~\ref{figure4}. The input state $|00\rangle$ does not pick any phase because the control atom is in Rydberg states which block the Rydberg excitation of the target atom. Third, apply the pulse shown in Fig.~\ref{figure1} for the control qubit with the pulse duration shown in Eq.~(\ref{sinpumping}) if the first theory is used, or the pulses shown in Figs.~\ref{figure4}(c) and~\ref{figure4}(d) if the second theory is used. These three steps will lead to an electronic C$_{\text{Z}}$ gate, i.e., an operation which maps the initial state
\begin{eqnarray}
 \sum_{\alpha,\beta\in\{0,1\}}a_{\alpha\beta}|\alpha\beta\rangle_{\text{e}}\otimes \sum_{\alpha,\beta\in\{0,1\}}b_{\alpha\beta}|\alpha\beta\rangle_{\text{n}},\label{initialstate2atomCZ}
\end{eqnarray}
to
\begin{eqnarray}
  &&\left[ -a_{00}|00\rangle_{\text{e}} -a_{01}|01\rangle_{\text{e}}- a_{10}|10\rangle_{\text{e}}+ a_{11}|11\rangle_{\text{e}}\right]\nonumber\\
  &&~~\otimes \sum_{\alpha,\beta\in\{0,1\}}b_{\alpha\beta}|\alpha\beta\rangle_{\text{n}},\label{initialstate2atomCZ2}
\end{eqnarray}
where $\sum_{\alpha,\beta\in\{0,1\}} |a_{\alpha\beta}|^2=\sum_{\alpha,\beta\in\{0,1\}} |b_{\alpha\beta}|^2=1$.

The fidelity to map Eq.~(\ref{initialstate2atomCZ}) to Eq.~(\ref{initialstate2atomCZ2}) is intrinsically limited by three factors. First, there is an intrinsic rotation error in the Rydberg excitation and deexcitation, as shown in Fig.~\ref{comparison2} for the first theory. Second, the Rydberg-state decay will lead to errors because there is time for the atomic state to be in the Rydberg states. Third, the blockade interaction is finite, which results in an imperfect blockade. The intrinsic rotation error is $1-\mathcal{F}$, where $\mathcal{F}$ is an average fidelity given by~\cite{Pedersen2007}
\begin{eqnarray}
\mathcal{F} &=&\left[  |\text{Tr}(U^\dag \mathscr{U})|^2 + \text{Tr}(U^\dag \mathscr{U}\mathscr{U}^\dag U ) \right]/272.\nonumber
\end{eqnarray}
Here, $\mathscr{U}$ is the actual gate matrix evaluated by using the unitary dynamics with the Rydberg-state decay ignored, and
$U$, given by,
\begin{eqnarray}
   \left(
  \begin{array}{cccc}
    -\openone_{12}&0\\
    0 &\openone_{4}
    \end{array}
  \right),  \label{Gatematrix}
\end{eqnarray}
is the ideal gate matrix in the ordered basis
\begin{eqnarray}
 &&|00\rangle_{\text{e}}\otimes\{|00\rangle_{\text{n}}, |01\rangle_{\text{n}}, |10\rangle_{\text{n}}, |11\rangle_{\text{n}}\},\nonumber\\
  && |01\rangle_{\text{e}}\otimes\{|00\rangle_{\text{n}}, |01\rangle_{\text{n}}, |10\rangle_{\text{n}}, |11\rangle_{\text{n}}\}, \nonumber\\
  &&|10\rangle_{\text{e}}\otimes\{|00\rangle_{\text{n}}, |01\rangle_{\text{n}}, |10\rangle_{\text{n}}, |11\rangle_{\text{n}}\},\nonumber\\
  &&|11\rangle_{\text{e}}\otimes\{|00\rangle_{\text{n}}, |01\rangle_{\text{n}}, |10\rangle_{\text{n}}, |11\rangle_{\text{n}}\}, \label{basisCZelectron}
\end{eqnarray}
where $\openone_{4}$ and $\openone_{12}$ are the $4\times4$ and $12\times12$ identity matrices, respectively. In order to evaluate the intrinsic rotation error, we assume that the Rydberg interaction is large enough and leave the blockade error to be analyzed separately when $V$ is finite~\cite{Saffman2005}. 

Because the method in Sec.~\ref{subsection02B02} is based on the theory shown in Ref.~\cite{Shi2021}, the fidelity analysis could be easily done following Ref.~\cite{Shi2021}. So, we will use the theory studied in Sec.~\ref{subsection02B01} to examine the intrinsic gate error. We employ Eqs.~(\ref{RydbergPumptheory01}) and~(\ref{RydbergPumptheory02}) to simulate the time dynamics for the states. During the first pulse, lasers with Hamiltonians $\hat{H}_1(t)$ in Eq.~(\ref{RydbergPumptheory01}) and $\hat{H}_2(t)$ in Eq.~(\ref{RydbergPumptheory02}) are applied for the control atoms with a duration $T_{\text{p1}}=$acos$[1-\pi\delta/(2\kappa_0)]/\delta\approx 1.30\times\frac{2\pi}{2\kappa_0}$ determined by Eq.~(\ref{sinpumping}). For the second pulse, lasers with Hamiltonians $\hat{H}_1(t-T_{\text{p1}})$ in Eq.~(\ref{RydbergPumptheory01}) and $\hat{H}_2(t-T_{\text{p1}})$ in Eq.~(\ref{RydbergPumptheory02}) are applied for the target atoms with a duration $T_{\text{p2}}=$acos$[1-\pi\delta/\kappa_0]/\delta\approx 1.89\times\frac{2\pi}{2\kappa_0}$, where $T_{\text{p2}}\neq 2T_{\text{p1}}$. During the third pulse, the same type of laser fields used in the first pulse are used, i.e., lasers with Hamiltonians $\hat{H}_1(t-T_{\text{p1}}-T_{\text{p2}})$ in Eq.~(\ref{RydbergPumptheory01}) and $\hat{H}_2(t-T_{\text{p1}}-T_{\text{p2}})$ in Eq.~(\ref{RydbergPumptheory02}) are applied for the control atoms with a duration $T_{\text{p1}}$. With the condition $\eta=1$ and $(\Delta,~\delta)/(2\kappa_0) = (10,~0.1)$, we numerically found that the intrinsic rotation error is $E_{\text{ro}}= 3.74\times10^{-4}$.
~The time for the atom to be in the Rydberg state averaged over the sixteen states in Eq.~(\ref{basisCZelectron}) is $T_{\text{Ryd}}=1.55\times\frac{2\pi}{2\kappa_0}$ as numerically evaluated, which leads to a decay error $E_{\text{decay}}=T_{\text{Ryd}}/\tau\approx3.35\times10^{-3}$ if we adopt the parameters from Ref.~\cite{Shi2021} with $2\kappa_0=2\pi\times1.4$~MHz and $\tau = 330~\mu$s. Finally, there will be a blockade error $ (2\kappa_0/V)^2$~\cite{Saffman2005} for each of the four states on the first row of Eq.~(\ref{basisCZelectron}). With $V=2\pi\times47$~MHz~\cite{Shi2021}, the error due to the blockade leakage for the gate will be $E_{\text{bl}}= 2.2\times10^{-4}$. So, the intrinsic gate fidelity for the electronic C$_{\text{Z}}$ operation described by Eq.~(\ref{initialstate2atomCZ}) and Eq.~(\ref{initialstate2atomCZ2}) is $1-E_{\text{ro}}-E_{\text{decay}}-E_{\text{bl}}\approx99.61\%$ which is dominated by the Rydberg-state decay.

\begin{figure}
\includegraphics[width=3.0in]
{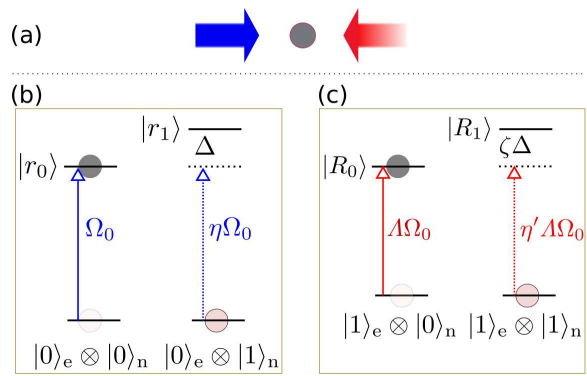}
\caption{ Rydberg excitation of $|0\rangle_{\text{e}}\otimes|0\rangle_{\text{n}}$ and $|1\rangle_{\text{e}}\otimes|0\rangle_{\text{n}}$. (a) An atom is excited by two sets of laser fields, one resonant with the transition from $|0\rangle_{\text{e}}\otimes|0\rangle_{\text{n}}$ to the Rydberg state $|r_0\rangle$, and the other is resonant with the transition from $|1\rangle_{\text{e}}\otimes|0\rangle_{\text{n}}$ to the Rydberg state $|R_0\rangle$. (b) There is a Zeeman shift in the Rydberg states (which is much larger than that in the ground or clock state), so the state  $|0\rangle_{\text{e}}\otimes|1\rangle_{\text{n}}$ is excited to a Rydberg state $|r_1\rangle$ with a detuning $\Delta$. (c) The Rydberg Rabi frequency is $\varLambda\Omega_0$ for the transition $|1\rangle_{\text{e}}\otimes|0\rangle_{\text{n}}\leftrightarrow|R_0\rangle$ where $\varLambda$ is tunable via adjustment of the strength of the laser fields. The other ground state $|1\rangle_{\text{e}}\otimes|1\rangle_{\text{n}}$ is excited to $|R_1\rangle$ with a detuning $\zeta\Delta$. Here, because there is some difference between the g factors of the ground and clock states, a factor $\zeta$ arises which is positive since the detuning is mainly determined by the electron g factors of the Rydberg states. The factor $\eta~(\eta')$ arises from angular momentum selection rules because the two nuclear spin states have different values of $m_I$. \label{NuclearPumping} }
\end{figure}

 \section{C$_{\text{Z}}$ gates with nuclear spin qubits}\label{section03}
HE in neutral atoms for our purpose is to entangle not only the electronic qubits, but also the nuclear spin qubits. We analyze methods to realize nuclear-spin C$_{\text{Z}}$ operation in this section. As in Sec.~\ref{section02}, we first outline methods to excite nuclear spin qubit states to Rydberg states. 

\subsection{Rydberg excitation of both electronic qubit states for a nuclear spin qubit state}\label{section03A}
As studied around Eq.~(\ref{state01}) in Sec.~\ref{section02A} where both nuclear spin states shall be excited for a certain electronic qubit state, here, when we want to excite the nuclear spin state $|0\rangle$ to Rydberg states, we actually need to map the state
\begin{eqnarray}
(\cos\theta|0\rangle_{\text{e}}+\sin\theta |1\rangle_{\text{e}} )\otimes|0\rangle_{\text{n}} \label{state01nuclear}
\end{eqnarray}
to a superposition of different Rydberg eigenstates
\begin{eqnarray}
\cos\theta|r_0\rangle+\sin\theta |R_0\rangle ,\label{state02nuclear}
\end{eqnarray}
where $\theta$ is a real variable~(here we ignore a relative phase between the two state components for brevity), $|r_{0}\rangle$ denotes the Rydberg state as used in Sec.~\ref{section02}, while $|R_0\rangle$ is a Rydberg state of a different principal quantum number.

To achieve Rydberg excitation of both electronic qubit states of a certain nuclear spin state, e.g., $|0\rangle$, we consider the laser configuration in Fig.~\ref{NuclearPumping}. Whether the theory in Sec.~\ref{subsection02B01} or the theory in Sec.~\ref{subsection02B02} is used, the Hamiltonian for the state $|0\rangle_{\text{e}}\otimes|0\rangle_{\text{n}}$ is 
\begin{eqnarray}
  \hat{H}_0&=& \frac{1}{2}   \left(
\begin{array}{cc}
    0& \Omega_0 \\
\Omega_0 &0\end{array}
  \right) \label{RydbergPumptheory01nuclear}
\end{eqnarray}
with the basis $\{|r_{0}\rangle,~|0\rangle_{\text{e}}\otimes|0\rangle_{\text{n}}\}$, and the Hamiltonian for the state $|1\rangle_{\text{e}}\otimes|0\rangle_{\text{n}}$ is 
\begin{eqnarray}
  \hat{H}_0'&=& \frac{1}{2}   \left(
\begin{array}{cc}
    0& \varLambda \Omega_0 \\
\varLambda\Omega_0 &0\end{array}
  \right) \label{RydbergPumptheory01nuclear2}
\end{eqnarray}
with the basis $\{|R_{0}\rangle,~|1\rangle_{\text{e}}\otimes|0\rangle_{\text{n}}\}$. Because the comparable magnitudes of the Rabi frequency and the Zeeman shift in the Rydberg states, the state $|0\rangle_{\text{e}}\otimes|1\rangle_{\text{n}}$ is excited off resonantly with the Hamiltonian 
\begin{eqnarray}
  \hat{H}_1&=& \frac{1}{2}   \left(
\begin{array}{cc}
    0& \eta \Omega_0 e^{i\Delta t}\\
  \eta   \Omega_0 e^{-i\Delta t}&0\end{array}
  \right) \label{RydbergPumptheory02nuclear}
\end{eqnarray}
with basis $\{|r_{1}\rangle,~|0\rangle_{\text{e}}\otimes|1\rangle_{\text{n}}\}$. Similarly, there is an off-resonant excitation of the state $|1\rangle_{\text{e}}\otimes|1\rangle_{\text{n}}$ with the Hamiltonian 
\begin{eqnarray}
  \hat{H}_1'&=& \frac{1}{2}   \left(
\begin{array}{cc}
    0& \eta'\varLambda  \Omega_0 e^{i\zeta\Delta t}\\
  \eta'\varLambda   \Omega_0 e^{-i\zeta\Delta t}&0\end{array}
  \right) \label{RydbergPumptheory02nuclear2}
\end{eqnarray}
with the basis $\{|R_{1}\rangle,~|1\rangle_{\text{e}}\otimes|1\rangle_{\text{n}}\}$.
Here, the detunings and dipole coupling matrix elements are different for the two transitions from the the electronic $|0\rangle$ and $|1\rangle$ states, which is incorporated in the factors $\zeta, \eta$, and $\eta'$ which are fixed variables~(once the polarization of the fields is fixed) because they are determined by the atomic states, while $\varLambda$ can be varied by adjustment of the strength of the laser fields.

\subsubsection{Rydberg excitation with sinusoidal pulses}\label{section03A01}
If the first theory, i.e., the theory in Sec.~\ref{subsection02B01}, is used, the Rabi frequencies $\Omega_0$ in Eq.~(\ref{RydbergPumptheory01nuclear}) and $\varLambda \Omega_0$ in Eq.~(\ref{RydbergPumptheory01nuclear2}) are both given by $2i\kappa_{0(1)}\sin(\delta t)$ as in Eq.~(\ref{RydbergPumptheory03}). Then, the Rydberg excitations of $|0\rangle_{\text{e}}\otimes|0\rangle_{\text{n}}$ and $|1\rangle_{\text{e}}\otimes|0\rangle_{\text{n}}$ have the same duration, although it is not a necessary condition, nor do we need to start the excitation of the two electronic states simultaneously. This is because the two electronic qubit states, namely, the ground state and the metastable clock state, have a frequency separation of hundreds of THz which is sufficiently large to suppress unwanted excitation of the non-targeted qubit state. When only one electronic qubit state is excited, the other electronic qubit state is not influenced. The benefit to have the same duration is that it can be easier to use pulse pickers to simultaneously switch the fields for exciting $|0\rangle_{\text{e}}\otimes|0\rangle_{\text{n}}$ and $|1\rangle_{\text{e}}\otimes|0\rangle_{\text{n}}$.    

There will be phase shifts to $|0\rangle_{\text{e}}\otimes|1\rangle_{\text{n}}$ and $|1\rangle_{\text{e}}\otimes|1\rangle_{\text{n}}$ when $|0\rangle_{\text{e}}\otimes|0\rangle_{\text{n}}$ and $|1\rangle_{\text{e}}\otimes|0\rangle_{\text{n}}$ are resonantly excited by the Hamiltonians in Eqs.~(\ref{RydbergPumptheory01nuclear}) and~(\ref{RydbergPumptheory01nuclear2}). By using the Hamiltonians in Eq.~(\ref{RydbergPumptheory02nuclear}), we numerically found that the phase shift to $|0\rangle_{\text{e}}\otimes|1\rangle_{\text{n}}$ is $\varphi_1\approx0.0395$ for a pulse duration $T_{\text{p1}}$=acos$[1-\pi\delta/(2\kappa_0)]/\delta$ when $\eta=1$ and $(\Delta,~\delta)/(2\kappa_0) = (10,~0.1)$. Although $|\zeta|,|\eta'|\approx1$ for the case of $^{171}$Yb analyzed in Ref.~\cite{Shi2021}, the phase shift to $|1\rangle_{\text{e}}\otimes|1\rangle_{\text{n}}$ will differ from $\delta\varphi$ if $|\zeta|,~|\eta'|\neq1$ when other isotopes are used. However, as will be shown in Sec.~\ref{section03A02}, the different phase shifts can be made equal by an extra phase compensation. Therefore, we will assume that the phase shifts are equal for the states $|0\rangle_{\text{e}}\otimes|1\rangle_{\text{n}}$ and $|1\rangle_{\text{e}}\otimes|1\rangle_{\text{n}}$.

\subsubsection{Rydberg excitation with rectangular pulses}\label{section03A02}
As shown in Sec.~\ref{subsection02B01}, the first theory needs a relatively large magnetic field, while the second theory in Sec.~\ref{subsection02B02} can be used with a small magnetic field. For the nuclear spin C$_{\text{Z}}$ operation with the second theory, the Hamiltonians for the relevant states are still given by Eqs.~(\ref{RydbergPumptheory01nuclear}),~(\ref{RydbergPumptheory01nuclear2}),~(\ref{RydbergPumptheory02nuclear}), and~(\ref{RydbergPumptheory02nuclear2}), but there are two differences compared to the first theory. First, quasi-rectangular pulses are used, i.e., $\Omega_0$ is constant when the laser pulse is sent, while the first theory requires shaped pulses. Second, there shall be a ``rational'' generalized Rabi frequency as specified below Eq.~(\ref{v0pmvdwI}), namely,  
\begin{eqnarray}
\sqrt{\eta^2\Omega_{0}^2+\Delta^2}&=&2N\Omega_0,\nonumber\\
\sqrt{(\eta')^2\varLambda^2\Omega_{0}^2+\zeta^2\Delta^2}&=&2N'\varLambda \Omega_0,\label{rabiratio}
\end{eqnarray}
where $N$ and $N'$ are positive integers. The above equations mean that if we choose the ratio of Rabi frequencies to be
\begin{eqnarray}
\varLambda &=& \frac{|\zeta\Delta|}{\sqrt{\Delta^2+\Omega_0^2[\eta^2-(\eta')^2]}},\label{etacondition}
\end{eqnarray}
the two equations in Eq.~(\ref{rabiratio}) can be satisfied with $N=N'$, which is feasible since $\Delta$ and the strength of the laser fields can be adjusted while the parameters $\zeta,~\eta$, and $\eta'$ are determined by the configuration of atomic levels. Because the $\pi$ pulse durations for Rydberg excitation in Fig.~\ref{NuclearPumping}(b) and~\ref{NuclearPumping}(c) are $\frac{\pi}{\Omega_0}$ and $\frac{\pi}{\varLambda\Omega_0}$, respectively, one can use the condition in Eq.~(\ref{etacondition}) to show that both the transition 
\begin{eqnarray}
|0\rangle_{\text{e}}\otimes|1\rangle_{\text{n}} \rightarrow e^{i\varphi_2} |0\rangle_{\text{e}}\otimes|1\rangle_{\text{n}}
\end{eqnarray}
with detuning $\Delta$ and the transition 
\begin{eqnarray}
|1\rangle_{\text{e}}\otimes|1\rangle_{\text{n}} \rightarrow e^{i\varphi_2'} |1\rangle_{\text{e}}\otimes|1\rangle_{\text{n}}
\end{eqnarray}
with detuning $\zeta\Delta$ acquire phases given by,
\begin{eqnarray}
\varphi_2 &=& -\left(N+\frac{\Delta}{2\Omega_{0}}\right)\pi,\nonumber\\
 \varphi_2'&=&-\left(N+\frac{\zeta\Delta}{2\varLambda\Omega_{0}}\right)\pi.\label{phasevarphiNuclear}
\end{eqnarray}
The above equation means that it is necessary to have extra phase compensation so as to avoid extra entanglement in the electronic states. This can be done by using highly detuned lasers tuned near to, e.g., $|1\rangle_{\text{e}}\otimes|1\rangle_{\text{n}}$, i.e., the other state $|1\rangle_{\text{e}}\otimes|0\rangle_{\text{n}}$ does not acquire any phase. Then, by exciting the state $|1\rangle_{\text{e}}\otimes|1\rangle_{\text{n}}$ with a Rabi frequency significantly smaller than the detuning to it, one can add an appropriate phase shift to it so that 
\begin{eqnarray}
 \varphi_2'&\rightarrow&\varphi_2.\label{primetovarphi}
\end{eqnarray}
Here, the fields may also excite the state $|1\rangle_{\text{e}}\otimes|0\rangle_{\text{n}}$ because it is almost degenerate with $|1\rangle_{\text{e}}\otimes|1\rangle_{\text{n}}$. However, the two nuclear spin states can be chosen in a way that $|0\rangle_{\text{n}}$ has the maximal nuclear spin projection along the external magnetic field, then using circularly polarized fields can avoid the excitation of $|1\rangle_{\text{e}}\otimes|0\rangle_{\text{n}}$ when $|1\rangle_{\text{e}}\otimes|1\rangle_{\text{n}}$ is excited; see Fig.~1(b) of Ref.~\cite{Shi2021} as an example. Here, it is necessary to choose the appropriate types of detuning since a blue~(red) detuning gives a tiny negative~(positive) phase shift for each detuned Rabi cycle~(with the understanding that a phase equal to an integer times $2\pi$ is trivial). A detailed analysis about such type of phase compensation can be found on page 6 of Ref.~\cite{Shi2021}. With the phase change in Eq.~(\ref{primetovarphi}), the Rydberg excitation changes the initial input state of the control qubit  
\begin{eqnarray}
  &&(\cos\theta|0\rangle_{\text{e}}+\sin\theta |1\rangle_{\text{e}} )\otimes(\cos\phi|0\rangle_{\text{n}}+\sin\phi |1\rangle_{\text{n}}),
 \label{state01nuclear02ini}
\end{eqnarray}
to
\begin{eqnarray}
&&-i \cos\phi (\cos\theta|r_0\rangle+\sin\theta |R_0\rangle)\nonumber\\
 &&+\sin\phi e^{i\varphi_2} (\cos\theta|0\rangle_{\text{e}}+\sin\theta |1\rangle_{\text{e}} )\otimes |1\rangle_{\text{n}},
 \label{state01nuclear02inter}
\end{eqnarray}
which can be followed by Rydberg excitation of the target atom so as to create a C$_{\text{Z}}$ gate in the nuclear spin state space.

When the nuclear spin state $|0\rangle$ is excited to Rydberg state and back at the end of a C$_{\text{Z}}$ gate pulse sequence, the state of the control qubit  
becomes
\begin{eqnarray}
&&- \cos\phi (\cos\theta|0\rangle_{\text{e}}+\sin\theta |1\rangle_{\text{e}} )\otimes |0\rangle_{\text{n}}\nonumber\\
 &&+\sin\phi e^{2i\varphi_2} (\cos\theta|0\rangle_{\text{e}}+\sin\theta |1\rangle_{\text{e}} )\otimes |1\rangle_{\text{n}},
 \label{state01nuclear02fin}
\end{eqnarray}
where the minus sign on the first line above arises from the $\pi+\pi$ pulses resonantly acting on the Rydberg excitation, and the phase $2\varphi_2$ on the second line arises from the detuned transition; it can be removed by extra pulses which will be shown in Sec.~\ref{section03B02}. Because only the nuclear spin $|0\rangle$ state is excited, the phase twist from the detuned transitions is quite different from that described in Fig.~\ref{figure4}. This is because in Fig.~\ref{figure4} both the two nuclear spin states should be excited to Rydberg states, while here we only need to excite one of the two nuclear spin states to Rydberg states. 

\subsection{C$_{\text{Z}}$ operation in the nuclear spin states and its fidelity}
\subsubsection{C$_{\text{Z}}$ gate with sinusoidal pulses}\label{section03B01}
The C$_{\text{Z}}$ gate sequence with the nuclear spin states can be realized as follows with the first theory, i.e, the theory in Sec.~\ref{subsection02B01}. First, apply the pulse sequence for the control atom shown in Fig.~\ref{NuclearPumping} with a duration $T$ specified by~[see Eq.~(\ref{sinpumping}) and explanations around it]
\begin{eqnarray}
\frac{\kappa_0[1-\cos(T\delta )]}{\delta}=\frac{\pi}{2},\label{pidurationtheory1Nuclearspin}
\end{eqnarray}
where $\Omega_0=2i\kappa_0\sin(\delta t)$ and $\varLambda=1$. Second, apply the laser fields for the target atom as in Fig.~\ref{NuclearPumping} for a duration $T$ with $\Omega_0=2i\kappa_0\sin(\delta t)$ and $\varLambda=1$. Third, repeat the second step, but with $\Omega_0=2i\kappa_0\sin(\delta t)e^{-i2\varphi_1}$ and $\varLambda=1$; here the relative phase $2\varphi_1$ for the Rabi frequencies in the third step is used to induce an extra phase shift to $|10\rangle_{\text{n}}$ because otherwise the final phases in $|10\rangle_{\text{n}}$ and $|11\rangle_{\text{n}}$ will be different. Fourth, apply the same pulse sequence in the first step so as to deexcite the Rydberg states of the control qubit. As a result of these four steps, a C$_{\text{Z}}$-like gate is realized which maps the initial state 
\begin{eqnarray}
 \sum_{\alpha,\beta\in\{0,1\}}a_{\alpha\beta}|\alpha\beta\rangle_{\text{e}}\otimes \sum_{\alpha,\beta\in\{0,1\}}b_{\alpha\beta}|\alpha\beta\rangle_{\text{n}},\label{initialstate2atomCZNulcear}
\end{eqnarray}
to
\begin{eqnarray}
  &&\sum_{\alpha,\beta\in\{0,1\}}a_{\alpha\beta}|\alpha\beta\rangle_{\text{e}}\otimes[ -b_{00}|00\rangle_{\text{n}} - b_{01}|01\rangle_{\text{n}}- b_{10}e^{4i\varphi_1}|10\rangle_{\text{n}}\nonumber\\
  &&~~~~~~~~+ b_{11}e^{4i\varphi_1}|11\rangle_{\text{n}} ] ,\label{CZnuclear1-withphase}
\end{eqnarray}
where $\varphi_1$ is the phase accumulation in the detuned Rabi oscillation as shown in Sec.~\ref{section03A01}. After a phase compensation strategy that will be shown in Sec.~\ref{section03B02}, the C$_{\text{Z}}$ operation emerges which leads to the output states, 
\begin{eqnarray}
  &&\sum_{\alpha,\beta\in\{0,1\}}a_{\alpha\beta}|\alpha\beta\rangle_{\text{e}}\otimes[ -b_{00}|00\rangle_{\text{n}} - b_{01}|01\rangle_{\text{n}}- b_{10}|10\rangle_{\text{n}}\nonumber\\
  &&~~~~~~~~+ b_{11}|11\rangle_{\text{n}} ] .\label{initialstate2atomCZ2Nulcear}
\end{eqnarray}

\subsubsection{C$_{\text{Z}}$ gate with rectangular pulses and phase compensation}\label{section03B02}
Here, we take the method in Sec.~\ref{subsection02B02} to discuss the phase compensation. As shown in Eq.~(\ref{state01nuclear02fin}), there will be an unwanted phase accumulation for the nuclear spin state. To solve this problem, we propose the following pulse sequence for realizing a C$_{\text{Z}}$-like gate and then describe methods to compensate the unwanted phase.

First, excite the control atom with the laser configuration shown in Fig.~\ref{NuclearPumping}. The pulse durations are $\pi/\Omega_0$ and $\pi/(\varLambda\Omega_0)$ for the $|0\rangle_{\text{e}}\otimes|0\rangle_{\text{n}}$ and $|1\rangle_{\text{e}}\otimes|0\rangle_{\text{n}}$ states, respectively. The two sets of laser fields in Figs.~\ref{NuclearPumping}(a) and~\ref{NuclearPumping}(b) do not need to start~(or end) at the same moment. This step excites the control atom to Rydberg states if the input nuclear spin states are $|00\rangle_{\text{n}}$ and $|01\rangle_{\text{n}}$.

Second, excite the target atom with laser fields shown in Fig.~\ref{NuclearPumping} with two pulses. In the first pulse, the Rabi frequency and pulse duration are $\Omega_0$ and $2\pi/\Omega_0$ for $|0\rangle_{\text{e}}\otimes|0\rangle_{\text{n}}$, and are $\varLambda\Omega_0$ and $2\pi/(\varLambda\Omega_0)$ for $|1\rangle_{\text{e}}\otimes|0\rangle_{\text{n}}$. In the second pulse, the Rabi frequency and pulse duration are $e^{-2i\varphi_2}\Omega_0$ and $2\pi/\Omega_0$ for exciting the input state $|0\rangle_{\text{e}}\otimes|0\rangle_{\text{n}}$, and are $e^{-2i\varphi_2}\varLambda\Omega_0$ and $2\pi/(\varLambda\Omega_0)$ for the input state $|1\rangle_{\text{e}}\otimes|0\rangle_{\text{n}}$. These two pulses resonantly excite the nuclear spin input state $|10\rangle_{\text{n}}$, which imprints a phase $\pi+2\varphi_2$ to it according to the picture of a standard resonant Rabi oscillation. However, the two pulses induce off-resonant excitation for the nuclear spin input state $|11\rangle_{\text{n}}$, imprinting a phase $2\varphi_2$ to it according to Eq.~(\ref{state01nuclear02fin}), where the phase twist $2\varphi_2$ is from the detuned Rabi cycles but not related with the phase of the Rabi frequencies.   
\begin{figure}
\includegraphics[width=3in]
{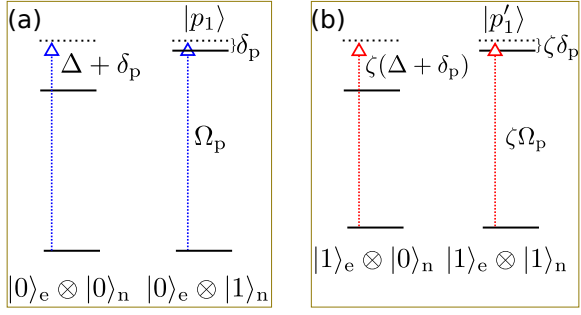}
\caption{Scheme to compensate the unwanted phase appeared in Eq.~(\ref{CZnuclear1-withphase}) or~(\ref{initialstate2atomCZ2Nulcear2}). A laser field is applied which is tuned near to the transition $|0\rangle_{\text{e}}\otimes |1\rangle_{\text{n}}\rightarrow|p_1\rangle$ with a blue detuning $|\delta_{\text{p}}|$, and similar for the transition $|1\rangle_{\text{e}}\otimes |1\rangle_{\text{n}}\rightarrow|p_1'\rangle$. Because $|\delta_{\text{p}}|\ll\Delta$, the states $|0(1)\rangle_{\text{e}}\otimes |0\rangle_{\text{n}}$ are highly detuned and do not pick up any phase error comparable to other errors.  \label{phasecompen2} }
\end{figure}

Third, apply the same pulses as in the first step for the control atom to restore the Rydberg state back to the ground state~(or the clock state). The first and third steps resonantly excite the nuclear spin input states $|00\rangle_{\text{n}}$ and $|01\rangle_{\text{n}}$, giving them a $\pi$ phase shift according to the standard picture of a resonant Rabi oscillation, and meanwhile results in a phase $2\varphi_2$ in the input states $|10\rangle_{\text{n}}$ and $|11\rangle_{\text{n}}$ according to Eq.~(\ref{state01nuclear02fin}). Therefore, the initial state in Eq.~(\ref{initialstate2atomCZNulcear}) is mapped to
\begin{eqnarray}
  &&\sum_{\alpha,\beta\in\{0,1\}}a_{\alpha\beta}|\alpha\beta\rangle_{\text{e}}\otimes[ -b_{00}|00\rangle_{\text{n}} - b_{01}|01\rangle_{\text{n}}- b_{10}e^{4i\varphi_2}|10\rangle_{\text{n}}\nonumber\\
  &&~~~~~~~~+ b_{11}e^{4i\varphi_2}|11\rangle_{\text{n}} ], \label{initialstate2atomCZ2Nulcear2}
\end{eqnarray}
where one can see that there is an unwanted phase $4\varphi_2$ for the input nuclear spin states $|10\rangle_{\text{n}}$ and $|11\rangle_{\text{n}}$. To undo this phase, we can use a strategy outlined in Fig.~\ref{phasecompen2} for the control atom, but with laser fields tuned nearly resonant, with Rabi frequency $\Omega_{\text{p}}$ and detuning $\delta_{\text{p}}$, to the transition $|0\rangle_{\text{e}}\otimes |1\rangle_{\text{n}}\rightarrow|p_1\rangle$; similarly, laser fields are used with Rabi frequency $\zeta\Omega_{\text{p}}$ and detuning $\zeta\delta_{\text{p}}$ for the transition $|1\rangle_{\text{e}}\otimes |1\rangle_{\text{n}}\rightarrow|p_1'\rangle$, where $|p_{1}\rangle$ and $|p_{1}'\rangle$ can be chosen from a high-lying Rydberg state or some low-lying states that have long lifetimes. The detuning shall be much smaller than $\Delta$, i.e., $|\delta_{\text{p}}|\ll\Delta$, so that the other nuclear spin states $|0(1)\rangle_{\text{e}}\otimes |0\rangle_{\text{n}}$ do not acquire extra phase shift for they are largely detuned when the Rabi frequencies satisfy the condition $\Omega_{\text{p}}\ll|\delta_{\text{p}}|$. With the following excitation
\begin{eqnarray}
|0\rangle_{\text{e}}\otimes |1\rangle_{\text{n}}&\xleftrightarrow[\text{Blue detuned by }\delta_{\text{p}}]{\Omega_{\text{p}}} &|p_1\rangle,\nonumber\\
|1\rangle_{\text{e}}\otimes |1\rangle_{\text{n}}&\xleftrightarrow[\text{Blue detuned by }\zeta\delta_{\text{p}}]{\zeta\Omega_{\text{p}}} &|p_1'\rangle,\label{eq52}
\end{eqnarray}
for a duration of $t_{\text{pc}}$ and $t_{\text{pc}}/|\zeta|$~(``pc'' represents phase compensation), respectively, one can show that~\cite{Shi2021}
\begin{eqnarray}
|0\rangle_{\text{e}}\otimes |1\rangle_{\text{n}}&\rightarrow&e^{i\varTheta t_{\text{pc}}\overline{\Omega_{\text{p}}}/(2\pi)} |0\rangle_{\text{e}}\otimes |1\rangle_{\text{n}}  ,\nonumber\\
|1\rangle_{\text{e}}\otimes |1\rangle_{\text{n}}&\rightarrow&e^{i\varTheta t_{\text{pc}}\overline{\Omega_{\text{p}}}/(2\pi)} |1\rangle_{\text{e}}\otimes |1\rangle_{\text{n}}.
\end{eqnarray}
by using the theoretical methods shown in Eqs.~(\ref{RydbergPumptheory02rotate}) to~(\ref{v0pmvdwI}), where $t_{\text{pc}}\overline{\Omega_{\text{p}}}/(2\pi)$ should be an integer. Here, $\overline{\Omega_{\text{p}}}\equiv\sqrt{\Omega_{\text{p}}^2+\delta_{\text{p}}^2}$ and
\begin{eqnarray}
\varTheta &=& -(1.0+\delta_{\text{p}}/\overline{\Omega_{\text{p}}})\pi,\label{phasecompen0001Nuclear}
\end{eqnarray}
which is approximately given by
\begin{eqnarray}
\varTheta \approx\left(-2+\frac{\Omega_{\text{p}}^2}{2\delta_{\text{p}}^2}\right)\pi,
\end{eqnarray}
in the limit $|\Omega_{\text{p}}/\delta_{\text{p}}|\ll1$. 

To undo the unwanted phase in Eq.~(\ref{initialstate2atomCZ2Nulcear2}), we impose the condition
\begin{eqnarray}
-2\pi\mathbb{N} &=& 4\varphi_2 + \varTheta t_{\text{pc}}\overline{\Omega_{\text{p}}}/(2\pi)
\end{eqnarray}
where $\mathbb{N}$ is a positive integer, leading to
\begin{eqnarray}
  t_{\text{pc}}&\approx& \frac{8\delta_{\text{p}}^2(2\varphi_2+\pi\mathbb{N})}{(4\delta_{\text{p}}^2 -\Omega_{\text{p}}^2)\sqrt{\Omega_{\text{p}}^2+\delta_{\text{p}}^2}},\label{eq57}
\end{eqnarray}
which can be further approximately as $2(2\varphi_2+\pi\mathbb{N} ) / |\delta_{\text{p}}| = 2\pi(\mathbb{N}-2N-\Delta/\Omega_0)/|\delta_{\text{p}}| $. Choosing the smallest integer $\mathbb{N}$ that is larger than $2N+\Delta/\Omega_0$ leads to a duration $t_{\text{pc}}$ that is smaller than $2\pi/|\delta_{\text{p}}|$. Here, $2\pi/|\delta_{\text{p}}|$ can be on the order of $ms$ when we have a MHz-scale $\Delta$ in the condition of $|\delta_{\text{p}}|\ll\Delta$.

With the unwanted phase terms removed in Eq.~(\ref{initialstate2atomCZ2Nulcear2}), the final output of the nuclear-spin C$_{\text{Z}}$ gate will be Eq.~(\ref{initialstate2atomCZ2Nulcear}). The above procedure is also applicable to undo the phase $4\varphi_1$ in Sec.~\ref{section03B01}. Thus, by using either the first or the second theory, we can realize a C$_{\text{Z}}$ operation in the nuclear spin space. 

\subsubsection{Gate fidelity}\label{section03B03}
The fidelity for the nuclear spin C$_{\text{Z}}$ operation is limited by three intrinsic factors, the Rydberg-state decay, the blockade leakage, and the intrinsic rotation error. The ideal gate matrix is still given by Eq.~(\ref{Gatematrix}), but with the basis 
\begin{eqnarray}
 &&\{|00\rangle_{\text{e}}, |01\rangle_{\text{e}}, |10\rangle_{\text{e}}, |11\rangle_{\text{e}}\}\otimes|00\rangle_{\text{n}},\nonumber\\
  && \{|00\rangle_{\text{e}}, |01\rangle_{\text{e}}, |10\rangle_{\text{e}}, |11\rangle_{\text{e}}\}\otimes|01\rangle_{\text{n}}, \nonumber\\
  &&\{|00\rangle_{\text{e}}, |01\rangle_{\text{e}}, |10\rangle_{\text{e}}, |11\rangle_{\text{e}}\}\otimes|10\rangle_{\text{n}},\nonumber\\
  &&\{|00\rangle_{\text{e}}, |01\rangle_{\text{e}}, |10\rangle_{\text{e}}, |11\rangle_{\text{e}}\}\otimes|11\rangle_{\text{n}}. \label{basisCZnuclearspin}
\end{eqnarray}

As in Sec.~\ref{electronicCZ}, we employ the Rydberg-excitation theory studied in Sec.~\ref{subsection02B01} to examine the intrinsic gate error since the analysis by the method in Sec.~\ref{subsection02B02} could be easily done following Ref.~\cite{Shi2021}. In principle, the Rabi frequencies from the ultraviolet~(UV) laser fields for the clock states, as shown in the experiment of Ref.~\cite{Madjarov2020}, can be much larger than that for the ground states. But for brevity, we assume the condition $\varLambda=1$ and apply the pulses with durations specified by Eq.~(\ref{pidurationtheory1Nuclearspin}) where $ \Omega_{0} = 2i\kappa_{0(1)}\sin(\delta t)$ and $(\Delta,~\delta)/(2\kappa_0) = (10,~0.1)$ as in Sec.~\ref{subsection02B01}. Moreover, we assume $|\eta|=|\eta'|=1$ corresponding to the case of $^{171}$Yb and $\pi$ polarized laser fields~\cite{Shi2021}, and assume $\zeta=1$ since the detunings are mainly determined by the Zeeman shifts of Rydberg states. Hamiltonians in Eqs.~(\ref{RydbergPumptheory01nuclear}) and~(\ref{RydbergPumptheory01nuclear2}) are used, and the duration for the first or the four pulse on the control atom, and that for the second or third pulse on the target atom are $T$=acos$[1-\pi\delta/(2\kappa_0)]/\delta$. The unwanted transitions in the other nuclear spin states are governed by Hamiltonians in Eqs.~(\ref{RydbergPumptheory02nuclear}) and~(\ref{RydbergPumptheory02nuclear2}). The targeted state transform is given in Eq.~(\ref{CZnuclear1-withphase}); we assume that the phase compensation can work perfectly that maps Eq.~(\ref{CZnuclear1-withphase}) to Eq.~(\ref{initialstate2atomCZ2Nulcear}), which is done by adding a phase $-4\varphi_1$ to the output of the input states $|10\rangle_{\text{n}}$ and $|11\rangle_{\text{n}}$ in the numerical simulation. The numerical result for the intrinsic rotation error is $E_{\text{ro}}= 2.63\times10^{-3}$ which is mainly determined by the population loss in the input state $|11\rangle_{\text{n}}$. The Rydberg superposition time is $T_{\text{Ryd}}=2.04\times\frac{2\pi}{2\kappa_0}$, so that the decay error is $E_{\text{decay}}=T_{\text{Ryd}}/\tau\approx4.42\times10^{-3}$ with $2\kappa_0=2\pi\times1.4$~MHz and $\tau = 330~\mu$s adopted from Sec.~\ref{subsection02B01}. The blockade error occurs for the four input states in the first line of Eq.~(\ref{basisCZnuclearspin}), so that we have $E_{\text{bl}}= 2.2\times10^{-4}$. The intrinsic fidelity for the nuclear spin C$_{\text{Z}}$ operation is $1-E_{\text{ro}}-E_{\text{decay}}-E_{\text{bl}}\approx99.27\%$.

Combining the results in Sec.~\ref{electronicCZ} where a fidelity 99.61\% was shown for the electronic C$_{\text{Z}}$ operation, the final fidelity is $ 99.61\%\cdot99.27\%=98.88\% $ for realizing the C$_{\text{Z}}\otimes$C$_{\text{Z}}$ gate which can map the initial state 
\begin{eqnarray}
 \sum_{\alpha,\beta\in\{0,1\}}a_{\alpha\beta}|\alpha\beta\rangle_{\text{e}}\otimes \sum_{\alpha,\beta\in\{0,1\}}b_{\alpha\beta}|\alpha\beta\rangle_{\text{n}},\label{initialstate2atomCZ-HE}
\end{eqnarray}
to
\begin{eqnarray}
  &&\left[ a_{00}|00\rangle_{\text{e}} +a_{01}|01\rangle_{\text{e}}+ a_{10}|10\rangle_{\text{e}}- a_{11}|11\rangle_{\text{e}}\right]\nonumber\\
  &&\otimes[ b_{00}|00\rangle_{\text{n}} +b_{01}|01\rangle_{\text{n}}+b_{10}|10\rangle_{\text{n}}- b_{11}|11\rangle_{\text{n}} ]\nonumber  
\end{eqnarray}
where $\sum_{\alpha,\beta\in\{0,1\}} |a_{\alpha\beta}|^2=\sum_{\alpha,\beta\in\{0,1\}} |b_{\alpha\beta}|^2=1$.

Note that the fidelity shown above does not mean that the C$_{\text{Z}}\otimes$C$_{\text{Z}}$ gate can not obtain higher fidelities. In the above estimate, the fidelity is small for the nuclear spin C$_{\text{Z}}$ operation for we have assumed $\Delta/(2\kappa_0) = 10$. But if $\Delta/(2\kappa_0) = 20$~(other parameters are the same), i.e., if the magnetic field is doubled compared to the condition above, the intrinsic rotation error in the nuclear spin C$_{\text{Z}}$ operation shrinks to $E_{\text{ro}}= 6.64\times10^{-4}$, leading to a fidelity $99.47\%$ for the nuclear spin C$_{\text{Z}}$ operation. Moreover, we have not used any optimization procedure. In principle, high-fidelity Rydberg blockade gates can be designed by using shaped pulses with appropriate time dependence in the amplitude or phase of the laser fields~\cite{Theis2016,Saffman2020,Guo2020,Li2021}.   

\section{Experimental consideration}\label{section04}
The practical feasibility of coherent Rydberg excitation is vital to the theory of this article, where two factors are of particular relevance: Rydberg excitation, and a MHz-scale Zeeman shift in the Rydberg state with a weak magnetic field. Below, we analyze two isotopes, $^{87}$Sr and $^{171}$Yb. One thing is common for these two elements, i.e., the hyperfine interaction mixes the singlet and triplet wavefunctions in their Rydberg states. Below, we first analyze $^{87}$Sr and briefly mention $^{171}$Yb for it was studied in Ref.~\cite{Shi2021} in detail. 


\subsection{$^{87}$Sr}
We first study strontium, an element extensively studied in experiments~\cite{Ido2003,Ye2013,Gaul2016,Winchester2017,Cooper2018,Covey2019,Norcia2018,Ding2018,Teixeira2020,Madjarov2020,Qiao2021} and in theories~\cite{Werij1992,Boyd2007,Vaillant2012,Vaillant2014,Dunning2016,Robicheaux2019,Mukherjee2011,Robertson2021}, which show that it is possible to prepare Rydberg states. For the Rydberg blockade, we calculate that the van der Waals interaction coefficient for the $(5s70s)^1S_0$ state is $C_6/2\pi=-710$~GHz$\mu$m$^6$ with the quantum defects used in Refs.~\cite{Robicheaux2019,Vaillant2012}. One can choose states with principal quantum numbers near $n=63$ where F\"{o}rster resonance occurs~\cite{Robicheaux2019} if strong interactions are desired. 

The Rydberg excitation of the clock state $(5s5p)^3P_0$ involves a one-photon UV laser excitation which is easily achievable with a large Rabi frequency~\cite{Madjarov2020}. For the ground state, a two-photon Rydberg excitation can be used where an intermediate state shall be used. The dipole matrix element between the ground state and $|(5s5p)^1P_1\rangle$ is large, but the short lifetime about $5$~ns of $|(5s5p)^1P_1\rangle$ can spoil the coherence during the excitation. There are two useful intermediate states for a two-photon Rydberg excitation, the state $|(5s5p)^3P_1\rangle$, and the state $|(5s6p)^1P_1\rangle$. As estimated in Ref.~\cite{Shi2021}, the dipole matrix element between the ground state and $|(5s5p)^3P_1\rangle$ is about half of that between the ground state and $|(5s6p)^1P_1\rangle$, while the dipole coupling matrices between these two intermediate states and a Rydberg state are of similar magnitude. So, the two-photon Rydberg Rabi frequency via the intermediate state $|(5s5p)^3P_1\rangle$ will be smaller if there is an upper bound for the achievable laser powers at hand. This means that only very stable Rydberg states can be coherently excited via $|(5s5p)^3P_1\rangle$.

\begin{figure}
\includegraphics[width=3.2in]
{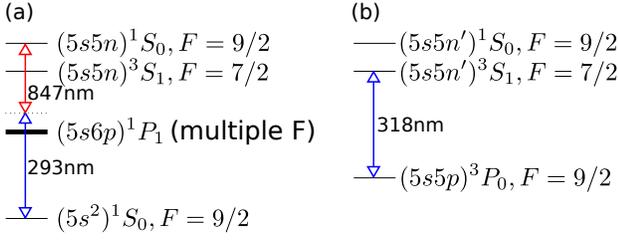}
\caption{(a) Energy level configuration for exciting a $(5sns)$ Rydberg state from the ground state of $^{87}$Sr. The intermediate state $|(5s6p)^1P_1\rangle$ is shown with a thick line for it has a hyperfine splitting of several tens of MHz which is negligible compared to a GHz-scale detuning at the intermediate state. The Rydberg state $|(5sns)^1S_0, F=9/2\rangle$ excited from the ground state is not purely singlet, but has some mixing of the triplet state $|(5sns)^3S_1, F=9/2\rangle$, shown in Sec.~\ref{sec04B2}, so that it has a g-factor dominated by the electron g-factor~(see Sec.~\ref{sec04B2} for detail). The nearest state to it is $|(5sns)^1S_0, F=7/2\rangle$ which is about $2\pi\times1.27$~GHz lower. (b) Energy level configuration of exciting $|(5sn's)^3S_1, F=7/2\rangle$ from the clock state by a one-photon transition, where $n'\neq n$. The Rydberg state nearest to it is $|(5sn's)^1S_0, F=9/2\rangle$~(which has some triplet component). The wavelengths were estimated by data from Ref.~\cite{Sansonetti2010}, quantum defects from Ref.~\cite{Ding2018}, and $n,n'\sim 70$.   \label{figure-SrEnergy} }
\end{figure}

Figure~\ref{figure-SrEnergy} shows the energy levels involved in the Rydberg excitation of the ground state via $|(5s6p)^1P_1\rangle$. The hyperfine interaction influences the level structures of the intermediate and Rydberg states which can be described by the frequency shift~\cite{Lurio1962,Arimondo1977},
\begin{eqnarray}
 E_{\text{hfs}} &=& A_{\text{hfs}} K + B_{\text{hfs}} \frac{\frac{3}{2}K(2K+1)-IJ(I+1)(J+1)}{2IJ(2I-1)(2J-1)},\nonumber\\ \label{Ehfs}
\end{eqnarray}
where $(I,J)=(9/2,1)$ here, and $K\equiv \mathbf{I}\cdot\mathbf{J}$ is expanded as
\begin{eqnarray}
 K&=& [F(F+1)-I(I+1)-J(J+1)]/2.
\end{eqnarray}
In Eq.~(\ref{Ehfs}), the first term is from the interaction of the nuclear magnetic moment and the magnetic field generated by the electrons, and the second term arises from the interaction between the electrons and the electric quadrupole moment of the nucleus.

The Rydberg excitation of the ground state in this work relies on an effective two-photon Rabi frequency. To show that it is possible to achieve a two-photon Rabi frequency, we need to analyze the level diagrams of the intermediate state and the Rydberg state.  

\subsubsection{Intermediate state}
We first analyze the hyperfine structure of the intermediate state. The quadrupole interaction exists only for states with $I, J\geq1$~\cite{Arimondo1977} which makes the level structures of $^{87}$Sr more complexed compared to that of $^{171}$Yb. The values of $A_{\text{hfs}} $ and $B_{\text{hfs}} $ can be measured by optical spectroscopy~\cite{Arimondo1977}. For the state $^{87}$Sr$|(5s5p)^1P_1\rangle$, we have $(A_{\text{hfs}} ,B_{\text{hfs}} )/2\pi= (-3.4,39)$~MHz~\cite{Kluge1974}, but we did not find experimental results of them for the state $|(5s6p)^1P_1\rangle$. However, one can theoretically estimate that the values of $A_{\text{hfs}}$ and $B_{\text{hfs}}$ are proportional to $l(l+1)\langle 1/r^3\rangle/[j(j+1)]$ and $(2j-1)\langle 1/r^3\rangle/(j+1)$, respectively~\cite{Lurio1962}, where $(l,j)=(1,1)$ for both $|(5s5p)^1P_1\rangle$ and $|(5s6p)^1P_1\rangle$, and $\langle 1/r^3\rangle$ is the expectation value of $1/r^3$ for the certain state. One can approximately have $\langle 1/r^3\rangle\propto 1/[{n^{\ast}}^3l(l+1/2)(l+1)]$~\cite{Schwartz1957}, where $n^{\ast}$ is the effective principal quantum number. With the approximation of $n^{\ast}=5$ and $6$ for the states $|(5s5p)^1P_1\rangle$ and $|(5s6p)^1P_1\rangle$, one can estimate that for $|(5s6p)^1P_1\rangle$ the hyperfine constants are $(A_{\text{hfs}} ,B_{\text{hfs}} )/2\pi= (-2.0,23)$~MHz.

\begin{figure}
\includegraphics[width=3in]
{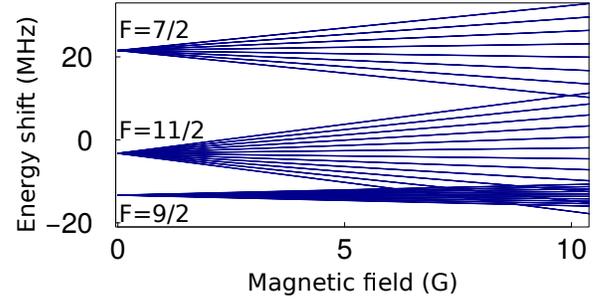}
\caption{Influence of the hyperfine and an external magnetic field $B$ on the energy level diagram (in reference to the unperturbed energy) for the state $^{87}$Sr$|(5s6p)^1P_1\rangle$. Presented here is an approximation in that $F$ is treated as a good quantum number which is valid only for small $B$. To convert to Joule, multiply by the Planck constant $h$.  \label{figure-ZeemanP} }
\end{figure}

With a magnetic field $B$ along $\mathbf{z}$, there is a Zeeman energy $E_{\text{Z}}=(g_sS_z\mu_{\text{B}}+g_lL_z\mu_{\text{B}}-g_II_z\mu_{\text{n}})B/\hbar$, where $g_s,g_l$, and $g_I$ are the electron spin, orbital, and the nuclear g factors, respectively, $S_z, L_z$, and $I_z$ are the z components of the electron spin, orbital, and nuclear spin angular momenta, respectively, and $\hbar$ is the reduced Planck constant. Here, $\mu_{\text{B}}$ is the Bohr magneton, $\mu_{\text{n}}=-1.0924\mu_{\text{N}}$ is the nuclear magnetic moment for $^{87}$Sr~\cite{Boyd2007}, and $\mu_{\text{N}}$ is the nuclear magneton. The energy levels of the atom can be numerically calculated by diagonalizing $E_{\text{hfs}}+E_{\text{Z}}$, but with $B<10$~G for $|(5s5p)^1P_1\rangle$, $F$ can be a good quantum number~\cite{Boyd2007}, and the energy shift~(divided by $\hbar$) of the atomic state is
\begin{eqnarray}
W   &=&  E_{\text{hfs}} +g_Fm_F\mu_{\text{B}}B/\hbar, \label{intermediateW}
\end{eqnarray}
where $g_F$ is the effective g-factor for the hyperfine substates. The hyperfine constants are of similar magnitudes for $|(5s6p)^1P_1\rangle$ and $|(5s5p)^1P_1\rangle$, so that we treat $F$ as a good quantum number with $B<10$~G for $|(5s6p)^1P_1\rangle$ and Eq.~(\ref{intermediateW}) is still useful here. With the approximation $g_L=1$ and neglecting diamagnetic correction which is rather tiny~\cite{Boyd2007}, we have $g_J\approx1$ for $|(5s6p)^1P_1\rangle$, so that $g_F\approx g_J[ F(F+1)-I(I+1)+ J(J+1)]/[2F(F+1)]$~\cite{DASteck}. The value of $W$ is shown in Fig.~\ref{figure-ZeemanP}, which shows that the energy separations between the hyperfine substates of different F are within $2\pi\times60$~MHz. When the single-photon detuning of the laser fields is about~\cite{Picken2018,Maller2015,Graham2019} or over~\cite{Isenhower2010,Zhang2010,Zeng2017} $2\pi\times1$~GHz, the different substates in Fig.~\ref{figure-ZeemanP} behave as one state for the energy difference between them is negligible compared to the the detuning. This means that a two-photon Rabi frequency between the ground and Rydberg states can be easily established for the case of $^{87}$Sr, which is in sharp contrast to the case of $^{171}$Yb analyzed in Ref.~\cite{Shi2021} where there is a GHz-scale hyperfine splitting in the intermediate state analyzed there.


\begin{figure}
\includegraphics[width=3.2in]
{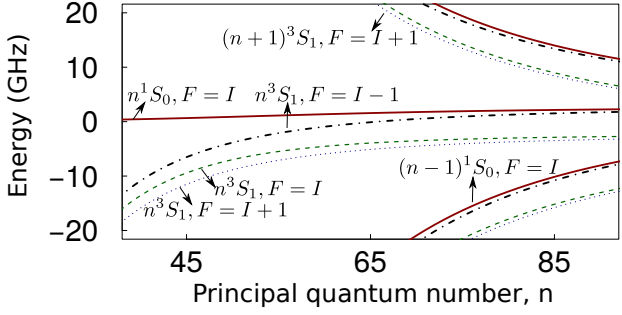}
\caption{Energy of a $(5sns)$-orbital Rydberg state of $^{87}$Sr in reference to the unperturbed energy of $(5sns)^1S_0$, i.e., the energy when neglecting the hyperfine interaction. For brevity, ``n'', instead of ``5sns'', labels the states. The quantum defects in~\cite{Ding2018} were used here. The topmost four curves are for the states with principal quantum number $n+1$, and the bottommost four curves are for the states with principal quantum number $n-1$. The states $|(5sns)^3S_1,F=I\pm1\rangle$ only obtain diagonal energy shifts in the presence of hyperfine interaction, but the states $|(5sns)^3S_1,F=I\rangle$ and $|(5sns)^1S_0,F=I\rangle$ are coupled with each other, so that the labels ``$n^3S_1,F=I$'' and ``$n^1S_0,F=I$'' beside the solid and dashed curves denote the main components in the mixed states. The energy is shown in units of GHz, to convert to Joule, multiply by $h$.  \label{figure-RydbergEnergy} }
\end{figure}
\subsubsection{Rydberg state}\label{sec04B2}
For $s$-orbital Rydberg states, the second term in Eq.~(\ref{Ehfs}) vanishes, but the hyperfine interaction is still there. The diagonal hyperfine coupling matrix elements are~\cite{Lurio1962,Ding2018}
\begin{eqnarray}
0&=&  \langle (5sns)^1S_0,F=I|A_{\text{hfs}}'\mathbf{I}\cdot\mathbf{J}|(5sns)^1S_0,F=I\rangle ,\nonumber\\
-\frac{A_{\text{hfs}}'}{2}&=& \langle (5sns)^3S_1,F=I|A_{\text{hfs}}'\mathbf{I}\cdot\mathbf{J}|(5sns)^3S_1,F=I\rangle ,\nonumber\\
\frac{A_{\text{hfs}}'I}{2}&=&  \langle (5sns)^3S_1,F=I+1|A_{\text{hfs}}'\mathbf{I}\cdot\mathbf{J}|(5sns)^3S_1,\nonumber\\
&& F=I+1\rangle , \label{hyperMatrix1}
\end{eqnarray}
and
\begin{eqnarray}
  -\frac{A_{\text{hfs}}'(I+1)}{2}&=& \langle (5sns)^3S_1,F=I-1|A_{\text{hfs}}'\mathbf{I}\cdot\mathbf{J}\nonumber\\
  &&|(5sns)^3S_1,F=I-1\rangle, \label{hyperMatrix2}
\end{eqnarray}
where $A_{\text{hfs}}'/2\pi\approx-1.0$~GHz is mainly due to the contact interaction between the $5s$ valence electron and the nucleus~\cite{Ding2018}. There is also an off-diagonal coupling
\begin{eqnarray}
  \frac{A_{\text{hfs}}'}{2}\sqrt{I(I+1)}\mathcal{O}_{n'n}&=& \langle (5sn's)^1S_0,F=I|A_{\text{hfs}}'\mathbf{I}\cdot\mathbf{J}\nonumber\\
  &&|(5sns)^3S_1,F=I\rangle, \label{hyperMatrix3}
\end{eqnarray}
where $\mathcal{O}_{n'n}\approx0.98$ for $n'=n$~\cite{Ding2018}. Because $\mathcal{O}_{n'n}<0.1$ for $n'\neq n$ and rapidly decreases when $|n'-n|$ increases, and also because the admixing of states with $|n'-n|>0$ results in small energy shift that can be incorporated in the practical experiment, we neglect terms with $n'\neq n$ in the theoretical analysis.

With the unperturbed energy of $(5sns)^1S_0$ as reference, Fig.~\ref{figure-RydbergEnergy} shows the energies of $|(5sns)^3S_1, F=I\pm1\rangle$ by the dotted and dash-dotted curves, respectively~(here we show the energy in reference to one $^{87}$Sr state for each n, in contrast to Fig.~2 of~\cite{Ding2018} which shows the difference of the energy of a $^{87}$Sr state and the energy of the $^{88}$Sr state of the same orbital). The solid and dashed curves in Fig.~\ref{figure-RydbergEnergy} are labeled with states that have the largest overlap of the eigenstates, and they denote states with both singlet and triplet components because Eq.~(\ref{hyperMatrix3}) shows that the states $|(5sns)^1S_0, F=I\rangle$ and $|(5sns)^3S_1, F=I\rangle$ are coupled by hyperfine interaction. Take $n=70$ as an example, the states labeled ``$n^1S_0, F=I$'' and ``$n^3S_1, F=I$'', separated by about $2\pi\times5.28$~GHz, have about 67\% and 33\% population in $|(5sns)^1S_0, F=I\rangle$, respectively. So, by choosing either of these two states for our theory, a g-factor dominated by the electron g-factor can lead to a MHz-scale Zeeman shift with a Gauss-scale magnetic field. Because the transition from the intermediate state $(5s6p)^1P_1$~(if it is purely singlet) to the state $|(5sns)^3S_1,F=I-1\rangle$ is spin forbidden, we can create HE via the state labeled ``$n^1S_0, F=I$'' for it has a larger state overlap with $|(5sns)^1S_0, F=I\rangle$, so that we can have a larger Rydberg Rabi frequency enabled by a large dipole matrix element between the Rydberg and intermediate states.

Population loss to nearby Rydberg states can be avoided. Take $n=70$ as an example, the state labeled `` $n^1S_0, F=I$'' for the solid curve is over the state $|(5sns)^3S_1, F=I-1\rangle$ by about $2\pi\times1.27$~GHz, and, more importantly, the transition from the intermediate state $(5s6p)^1P_1$ to the pure triplet state $|(5sns)^3S_1, F=I-1\rangle$ is spin forbidden; there can be a tiny mixing of the triplet wavefunction in $(5s6p)^1P_1$ as shown in~\cite{Boyd2007}, so, the Rabi frequency of the transition from $(5s6p)^1P_1$ to $|(5sns)^3S_1, F=I-1\rangle$ will be several orders of magnitude smaller than the GHz-scale detuning. This can suppress unwanted population loss. The next nearest state is the state labeled ``$n^3S_1, F=I$'', but it is about $2\pi\times5.28$~GHz below ``$n^1S_0, F=I$'', so that the population leakage to it can be neglected.

\subsection{$^{171}$Yb}\label{section04A}
We then briefly discuss $^{171}$Yb since there is a detailed analysis in Ref.~\cite{Shi2021}. The two nuclear spin states with $m_I=\pm1/2$ define a nuclear spin qubit, and the ground state $(6s^2)^1S_0$ and the clock state $(6s6p)^3P_0$ define an electronic qubit. The ground state has a pure singlet pairing in the two valence electrons, while there is a tiny singlet component mixed in the clock state~\cite{Boyd2007}. Without such a mixing, the lifetime of the clock state would be as long as the ground state. Because the mixing is tiny, the lifetime of the clock state is still comparable to that of the ground state.

The excitation from the clock state $(6s6p)^3P_0$ to a $^3S_1$ Rydberg states is easy to realize with UV lasers~\cite{Hankin2014,DeSalvo2016,Madjarov2020}, and the $^3S_1$ Rydberg state has a g-factor dominated by the electron g-factor so that a large Zeeman shift can appear with a weak magnetic field. However, it is questionable whether fast Rydberg excitation from the ground to Rydberg states can be achieved. We find that it is possible to use the $(6s6p)^3P_1$ state as the intermediate state for Rydberg excitation. As shown in Ref.~\cite{Budick1970}, the state $|(6s6p)^3P_1\rangle$ is actually given by $\beta|(6s6p)^1P_1^0\rangle + \alpha|(6s6p)^3P_1^0\rangle $ with $(\alpha,~\beta)=(0.991,~-0.133)$~\cite{Budick1970}, where the superscript $0$ denotes pure Russell-Saunders states. Because the transition from the singlet to triplet states is spin forbidden, the ground state can be coupled to the $(6s6p)^1P_1^0$ component in $(6s6p)^3P_1$. The dipole matrix element between the ground state and $(6s6p)^1P_1^0$ can be estimated by the Weisskopf-Wigner approximation~\cite{DASteck}, leading to a matrix element on the order of $ea_0$ where $e$ is the elementary charge and $a_0$ is the Bohr radius. With the mixing coefficient $|\beta|=0.133$, the dipole matrix element between the ground state and the intermediate state is on the order of $0.1ea_0$. For the upper transition from $|(6s6p)^3P_1\rangle$ to $(6sns)^3S_1\rangle$ with $n$ a large principal quantum number, we use the semi-classical analytical formula~\cite{Kaulakys1995} which was tested to be a useful approximation~\cite{Walker2008}. Then one can estimate a dipole coupling matrix element about $0.005ea_0$ with $n\sim70$~\cite{Covey2019pra}. The transition from the ground state to the $(6s6p)^3P_1$ state needs light of wavelength about $556$~nm~\cite{Pandey2009,Atkinson2019}, and its transition to the Rydberg state with $n\sim70$ needs radiation of wavelength $308.4$~nm~\cite{Shi2021}. In the experiments of~\cite{Higgins2017} lasers with wavelength in the range $304-309$~nm were used to excite Rydberg states of a strontium ion~(see also Refs.~\cite{Higgins2017prl,Zhang2020}). These UV lasers could be prepared by frequency doubling via the second-harmonic generation. There is a hyperfine splitting about $2\pi\times5.9$~GHz in the $|(6s6p)^3P_1\rangle$ state, which makes it necessary to numerically verify the possibility of a two-photon Rydberg excitation. This was done in Ref.~\cite{Shi2021}, which showed that it is possible to obtain an effective Rydberg Rabi frequency over $2\pi\times1$~MHz for the transition between the ground and Rydberg states via the $(6s6p)^3P_1$ intermediate state. The g-factor of the $(6sns)^3S_1$ state is dominated by the electron g-factor, so that a MHz-scale Zeeman shift can arise with a Gauss-scale magnetic field. The energy levels and the hyperfine couplings involved in the Rydberg excitation are shown in Ref.~\cite{Shi2021}. 

The Rydberg interaction can be large engouh for ytterbium, too. By quantum defects in Ref.~\cite{Lehec2018} and radial integration of Ref.~\cite{Kaulakys1995}, we calculate~\cite{Shi2014} that the van der Waals coefficient is $C_6/2\pi=32$~GHz$\mu$m$^6$ for, as an example, the ytterbium $(6s70s)^1S_0$ state if the atoms are along the quantization axis. However, the quantum defects for the $^3P_{0,1,2}$ Rydberg states of ytterbium are not available but they are required to calculate the interaction for atoms in the $^3S_1$ states, which should be much larger. This is because the interaction in $^3S_1$ atoms has nine transition channels, while two atoms in the $^1S_0$ state only have one transition channel. So, Rydberg blockade can take effect with our theories. 

\begin{figure}
\includegraphics[width=3.0in]
{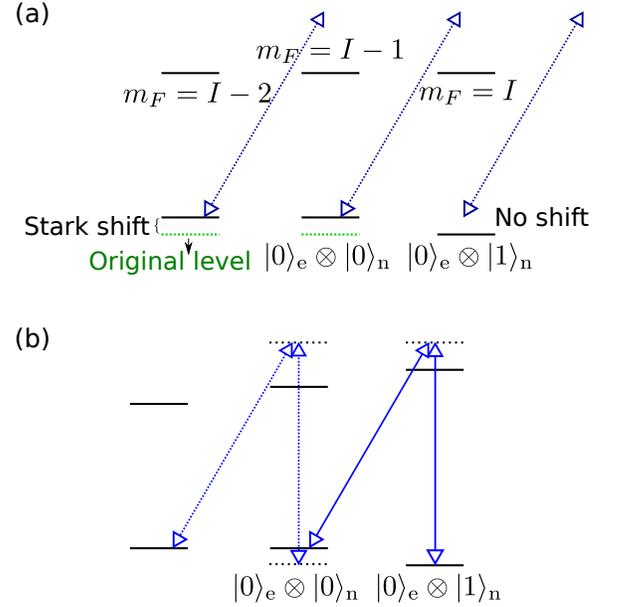}
\caption{Example of population transfer between the nuclear spin states in the ground state of $^{87}$Sr. (a) By exciting the ground state to a $^3P_0$ state~[such as $(5s7p)^3P_0$] with highly off-resonant, right-hand polarized laser fields, a MHz-scale Stark shift can appear for the ground Zeeman substates $m_F=I-1, I-2,\cdots$, but there is negligible shift for the state $m_F=I$ for there is no $m_F=I+1$ state in the $^3P_0$ state. (b) A two-photon transition between the two ground Zeeman substates $m_F=I$ and $m_F=I-1$, i.e., $|0\rangle_{\text{e}}\otimes|1\rangle_{\text{n}}$ and $|0\rangle_{\text{e}}\otimes|0\rangle_{\text{n}}$, is established by optical excitation of them via a highly off-resonant $p$-orbital state~[such as the state $(5s5p)^3P_1$]. Because the Stark shifts shown in (a), the transition between $|0\rangle_{\text{e}}\otimes|0\rangle_{\text{n}}$ and the ground Zeeman substate $m_F=I-2$ is highly detuned, so that there is no population leakage outside the qubit state space.  \label{figure-singleQubit} }
\end{figure}

\section{Single-qubit gates}\label{section05}
Single-qubit logic operations can proceed based on the methods shown above.

Single-qubit gates with the nuclear qubits are as follows. (i) To induce a phase shift like $|0\rangle_{\text{e}}\otimes|1\rangle_{\text{n}}\rightarrow e^{-2\varphi_2}|0\rangle_{\text{e}}\otimes|1\rangle_{\text{n}} $, we can use the highly off-resonant optical excitation shown in Fig.~\ref{phasecompen2}(a), where the contents around Eqs.~(\ref{eq52})-(\ref{eq57}) show that the phase-shift gate can complete within a time on the order of ms. To induce a phase shift to the state $|0\rangle_{\text{e}}\otimes|0\rangle_{\text{n}}$, the fields can be tuned near to the transition between $|0\rangle_{\text{e}}\otimes|0\rangle_{\text{n}}$ and the Rydberg state. (ii) To create a population transfer between $|0\rangle_{\text{e}}\otimes|0\rangle_{\text{n}}$ and $|0\rangle_{\text{e}}\otimes|1\rangle_{\text{n}}$, the method in Fig.~\ref{figure-singleQubit} can be used, where Fig.~\ref{figure-singleQubit}(a) shows that using Stark shift one can create energy shift to the Zeeman substates in the ground state. The two-photon Raman transition via the $p$-orbital state with $m_F=I$ in Fig.~\ref{figure-singleQubit}(b) can induce coherent population transfer between $|0\rangle_{\text{e}}\otimes|0\rangle_{\text{n}}$ and $|0\rangle_{\text{e}}\otimes|1\rangle_{\text{n}}$. But because of the Stark shifts in Fig.~\ref{figure-singleQubit}(a), the leakage from the state $|0\rangle_{\text{e}}\otimes|0\rangle_{\text{n}}$ to the Zeeman substate with $m_F=I-2$ is highly detuned so as to suppress population leakage out of the computational basis. (iii) The phase-shift gate and population transfer operation with $|1\rangle_{\text{e}}\otimes|1\rangle_{\text{n}}$ and $|1\rangle_{\text{e}}\otimes|1\rangle_{\text{n}}$ can proceed as in (i) and (ii), but with $s$ or $d$-orbital states instead of the $p$-orbital states in Fig.~\ref{phasecompen2} and Fig.~\ref{figure-singleQubit} for inducing phase shift or Raman transition. (iv) Combining (i) and (iii), one can create a phase shift to either of the two nuclear spin qubit states; combining (ii) and (iii), a population transfer between the two nuclear spin qubit states can occur.

Single-qubit phase gates with the electronic qubits can proceed as described by the contents in (i) and (iii) of the last paragraph. For population transfer between the electronic qubits, laser fields polarized along the quantization axis can induce transitions
\begin{eqnarray}
&&  |0\rangle_{\text{e}}\otimes|0\rangle_{\text{n}}\leftrightarrow |1\rangle_{\text{e}}\otimes|0\rangle_{\text{n}} ,\label{electronPop1}\\
&&  |0\rangle_{\text{e}}\otimes|1\rangle_{\text{n}}\leftrightarrow |1\rangle_{\text{e}}\otimes|1\rangle_{\text{n}} .\label{electronPop2}
\end{eqnarray}
Because of the hyperfine interaction that modifies the wavefunction of the clock state, there is a linear Zeeman shift which is about $\delta_{\text{s-p}}=B\times0.11m_F$~kHz$/$G for the $^1S_0-^3P_0$ transition~\cite{Boyd2007}. Then, if the transition in Eq.~(\ref{electronPop1}) is resonant, the transition in Eq.~(\ref{electronPop2}) is with a detuning $\delta_{\text{s-p}}(m_F=I)-\delta_{\text{s-p}}(m_F=I-1)$, and vice versa. So, to induce a population transfer for the same amount for both Eq.~(\ref{electronPop1}) and Eq.~(\ref{electronPop2}), the methods outlined in Sec.~\ref{section02B} with the specific strategy shown in Fig.~\ref{figure1} or Fig.~\ref{figure4} can be used, where the lower and upper states in Figs.~\ref{figure1} and~\ref{figure4} correspond to the states on the left and right sides of Eqs.~(\ref{electronPop1}) and~(\ref{electronPop2}), respectively.

\section{Cross-entanglement}\label{section06}
\subsection{Entanglement in one atom}\label{section06A}
It is possible to induce entanglement between the electronic and nuclear spin qubit states within one atom. In particle, we consider such a C$_{\text{Z}}$ operation that maps the state $( |0\rangle_{\text{e}}\otimes|0\rangle_{\text{n}} + |0\rangle_{\text{e}}\otimes|1\rangle_{\text{n}} +  |1\rangle_{\text{e}}\otimes|0\rangle_{\text{n}} +  |1\rangle_{\text{e}}\otimes|1\rangle_{\text{n}})/2$ to $( |0\rangle_{\text{e}}\otimes|0\rangle_{\text{n}} + |0\rangle_{\text{e}}\otimes|1\rangle_{\text{n}} +  |1\rangle_{\text{e}}\otimes|0\rangle_{\text{n}} -|1\rangle_{\text{e}}\otimes|1\rangle_{\text{n}})/2$. This operation is simply a phase shift operation for the state $|1\rangle_{\text{e}}\otimes|1\rangle_{\text{n}}$, which can be achieved by the strategy shown in the second paragraph of Sec.~\ref{section05}.

\subsection{Entanglement between two atoms}\label{section06B}
It is also possible to create entanglement between the electronic qubit states of one atom and the nuclear spin qubit states of another atom. We consider a C$_{\text{Z}}$ operation between the electron state $|1\rangle$ in the control atom and the nuclear spin state $|1\rangle$ in the target atom. To avoid confusion, we use subscripts c and t to denote states for the control and target atoms, respectively, then this gate maps the state
\begin{eqnarray}
 && ( |0\rangle_{\text{e}}\otimes|0\rangle_{\text{n}} + |0\rangle_{\text{e}}\otimes|1\rangle_{\text{n}} +  |1\rangle_{\text{e}}\otimes|0\rangle_{\text{n}} +  |1\rangle_{\text{e}}\otimes|1\rangle_{\text{n}})_{\text{c}}\nonumber\\
 && \otimes( |0\rangle_{\text{e}}\otimes|0\rangle_{\text{n}} + |0\rangle_{\text{e}}\otimes|1\rangle_{\text{n}} +  |1\rangle_{\text{e}}\otimes|0\rangle_{\text{n}} +  |1\rangle_{\text{e}}\otimes|1\rangle_{\text{n}})_{\text{t}}\nonumber\\ \label{crossEntangle1}
\end{eqnarray}
to 
\begin{eqnarray}
 && ( |0\rangle_{\text{e}}\otimes|0\rangle_{\text{n}} + |0\rangle_{\text{e}}\otimes|1\rangle_{\text{n}} )_{\text{c}}\nonumber\\
 && \otimes( |0\rangle_{\text{e}}\otimes|0\rangle_{\text{n}} + |0\rangle_{\text{e}}\otimes|1\rangle_{\text{n}} +  |1\rangle_{\text{e}}\otimes|0\rangle_{\text{n}} +  |1\rangle_{\text{e}}\otimes|1\rangle_{\text{n}})_{\text{t}}\nonumber\\
 && +(  |1\rangle_{\text{e}}\otimes|0\rangle_{\text{n}} +  |1\rangle_{\text{e}}\otimes|1\rangle_{\text{n}})_{\text{c}}\otimes( |0\rangle_{\text{e}}\otimes|0\rangle_{\text{n}}  +  |1\rangle_{\text{e}}\otimes|0\rangle_{\text{n}} )_{\text{t}}\nonumber\\  
 &&- (  |1\rangle_{\text{e}}\otimes|0\rangle_{\text{n}} +  |1\rangle_{\text{e}}\otimes|1\rangle_{\text{n}})_{\text{c}}\otimes(  |0\rangle_{\text{e}}\otimes|1\rangle_{\text{n}} +  |1\rangle_{\text{e}}\otimes|1\rangle_{\text{n}})_{\text{t}}. \nonumber\\ \label{crossEntangle2}
\end{eqnarray}

The above C$_{\text{Z}}$ operation can be realized as follows. (i) Use the strategy specified in the first paragraph of Sec.~\ref{electronicCZ} to excite the state $(|0\rangle_{\text{e}}\otimes|0\rangle_{\text{n}} +  |0\rangle_{\text{e}}\otimes|1\rangle_{\text{n}})_{\text{c}}$ to Rydberg states for the control atom. For example, the two step in Figs.~\ref{figure4}(a) and~\ref{figure4}(b) can realize the Rydberg excitation. (ii) Use either the method in Sec.~\ref{section03B01} or the method in Sec.~\ref{section03B02} to excite the state $(|0\rangle_{\text{e}}\otimes|1\rangle_{\text{n}} +  |1\rangle_{\text{e}}\otimes|1\rangle_{\text{n}})_{\text{t}}$ of the target atom to Rydberg states and back again. When there is no Rydberg blockade from the control atom, the state $(|0\rangle_{\text{e}}\otimes|1\rangle_{\text{n}} +  |1\rangle_{\text{e}}\otimes|1\rangle_{\text{n}})_{\text{t}}$ will pick up a $\pi$ phase during this step. (iii) Use similar laser excitation as used in step (i), but with $\pi$ phase different in the Rydberg Rabi frequencies. Take the Rydberg deexcitation shown in Fig.~\ref{figure4} as an example, the resonant Rabi frequencies should be $-e^{2i\varphi}\Omega_1$ and $-e^{2i\varphi}\Omega_0$ in Figs.~\ref{figure4}(c) and~\ref{figure4}(d), respectively. Then, there will be no phase twist to the state $(|0\rangle_{\text{e}}\otimes|0\rangle_{\text{n}} +  |0\rangle_{\text{e}}\otimes|1\rangle_{\text{n}})_{\text{c}}$. These three steps can realize the map from Eq.~(\ref{crossEntangle1}) to Eq.~(\ref{crossEntangle2}).   

\section{Conclusions}\label{section07}
We study hyperentanglement in divalent neutral atoms realized by exciting the ground and clock states of AEL atoms to Rydberg levels. Our theories take advantage of the fact that in the ground and clock states the electronic and nuclear degrees of freedom are decoupled. We show that without changing the states of the electronic qubits, nuclear spin qubits can be entangled between two atoms. On the other hand, without changing the states of the nuclear spin qubits, the electronic qubits can be entangled, too. Detailed analysis shows that a fidelity over 98\% can be achieved for realizing the C$_{\text{Z}}\otimes$C$_{\text{Z}}$ operation in the electronic and nuclear spin qubits of two atoms. The possibility to create hyperentanglement with neutral atoms sheds new light on the study of quantum control with neutral atoms.

\section*{ACKNOWLEDGMENTS}
The author thanks Yan Lu for useful discussions. This work is supported by the National Natural Science Foundation of China under Grants No. 12074300 and No. 11805146, the Natural Science Basic Research plan in Shaanxi Province of China under Grant No. 2020JM-189, and the Fundamental Research Funds for the Central Universities.

%


\end{document}